\documentclass[]{JFM-FLM_Au}
\usepackage{amsmath,lipsum}

\lefttitle{V. Gomez Herrera and S. Weady}
\righttitle{Journal of Fluid Mechanics}

\title{Instability of microbial droplets growing on viscous substrates}

\author{Vicente Gomez Herrera\aff{1,2} \and Scott Weady\aff{2}}

\affiliation{
\aff{1}Courant Institute of Mathematical Sciences, New York University, New York, NY 10012, USA
\aff{2}Center for Computational Biology, Flatiron Institute, New York, NY 10010, USA
}

\corresau{Scott Weady, \email{sweady@flatironinstitute.org}}

\newcommand{\eps}{\varepsilon}

\newcommand{\ft}{\hat}
\newcommand{\wft}{\widehat}

\renewcommand{\d}{{\rm d}}
\renewcommand{\i}{{\rm i}}

\newcommand{\bbR}{\mathbb{R}}

\newcommand{\grads}{\nabla_{\rm 2D}}
\newcommand{\laps}{\Delta_{\rm 2D}}
\newcommand{\gradv}{\nabla_{\rm 3D}}
\newcommand{\lapv}{\Delta_{\rm 3D}}

\newcommand{\be}{\boldsymbol{e}}
\newcommand{\bfv}{\boldsymbol{f}}

\newcommand{\bk}{\boldsymbol{k}}

\newcommand{\bn}{\boldsymbol{n}}

\newcommand{\br}{\boldsymbol{r}}

\newcommand{\bu}{\boldsymbol{u}}

\newcommand{\bx}{\boldsymbol{x}}
\newcommand{\by}{\boldsymbol{y}}

\newcommand{\bG}{\mathsfbi{G}}

\newcommand{\bI}{\mathsfbi{I}}

\newcommand{\bU}{\boldsymbol{U}}

\newcommand{\bX}{\boldsymbol{X}}

\newcommand{\bxi}{\boldsymbol{\xi}}

\newcommand{\bzero}{\boldsymbol{0}}
\newcommand{\bGamma}{\boldsymbol{\Gamma}}
\newcommand{\btheta}{\boldsymbol{\theta}}

\newcommand{\cB}{\mathcal{B}}
\newcommand{\cF}{\mathcal{F}}

\newcommand{\cI}{\mathcal{I}}
\newcommand{\cN}{\mathcal{N}}
\newcommand{\cS}{\mathcal{S}}
\newcommand{\cT}{\mathcal{T}}
\newcommand{\cV}{\mathcal{V}}

\newcommand{\cBeps}[1]{\delta\mathcal{B}_{D_{#1}}^m}
\newcommand{\cNeps}[1]{\delta\mathcal{N}_{D_{#1}}^m}
\newcommand{\cSeps}[1]{\delta\mathcal{S}_{D_{#1}}^m}
\newcommand{\cTeps}[1]{\delta\mathcal{T}_{D_{#1}}^m}
\newcommand{\cVeps}[1]{\delta\mathcal{V}_{D_{#1}}^m}

\newcommand{\feps}{f_{R,\eps}^m}
\newcommand{\Jeps}[1]{\delta {J}_{D_{#1}}^m}

\def\tbu{\tilde\bu}

\def\tnabla{\tilde\nabla}

\def\Pec{Pe}
\def\Ray{Ra}

\def\pg{p_g}
\def\bug{\bu_g}
\def\pb{p_b}
\def\bub{\bu_b}
\def\sigmab{\sigma_b}

\newcommand{\pzeta}{{\zeta^\prime}}

\def\Xint#1{\mathchoice
   {\XXint\displaystyle\textstyle{#1}}%
   {\XXint\textstyle\scriptstyle{#1}}%
   {\XXint\scriptstyle\scriptscriptstyle{#1}}%
   {\XXint\scriptscriptstyle\scriptscriptstyle{#1}}%
   \!\int}
\def\XXint#1#2#3{{\setbox0=\hbox{$#1{#2#3}{\int}$}
     \vcenter{\hbox{$#2#3$}}\kern-.5\wd0}}
\def\fpint{\Xint=}

\begin{document}
\maketitle

\begin{abstract}
We develop and analyze a model for a flat microbial droplet growing on the surface of a three-dimensional viscous fluid. The model describes growth-induced stresses at the fluid surface, density variations in the bulk due to nutrient consumption, and the resulting fluid flows that arise. We reformulate this free-boundary problem as a system of integro-differential equations defined solely on the microbial domain. From this formulation, we identify an axisymmetric solution corresponding to a radially expanding disk and analyze its morphological stability. We find that growth forces stabilize the axisymmetric solution while buoyancy forces destabilize it. We connect these findings to experimental observations.
\end{abstract}

\section{Introduction}

Microbial communities are strongly influenced by the environments they inhabit. For example, the growth and morphology of bacterial colonies and biofilms depend on substrate composition and rheology \citep{fei2020nonuniform,asp2022spreading,faiza2025substrate,gonzalez2025morphogenesis}, substrate topography controls motility and nutrient accessibility \citep{chang2015biofilm,gu2016escherichia,postek2024substrate}, and ambient fluid flows affect morphology and population dynamics \citep{perlekar2010population,pearce2019flow,benzi2022spatial}. These features, in many cases, arise from a feedback loop between growth, mechanics, and chemical sensing \citep{hallatschek2023proliferating}.

Microbial communities at or near fluid interfaces arise in a variety of culinary, industrial, and environmental contexts and present a particularly interesting case \citep{vaccari2017films}. Examples include kilometer-scale algal blooms that are mixed by turbulent oceanic flows \citep{abraham1998generation,hallegraeff2003harmful}, centimeter-scale pellicles that facilitate kombucha fermentation \citep{aung2024comprehensive}, and millimetric biofilms on oil droplets that can aid in bioremediation \citep{vaccari2015films,hickl2022tubulation, prasad2023alcanivorax}. Experiments of yeast colonies ({\em S. cerevisiae}) growing on viscous substrates found that metabolically generated flows, which arise from depletion of dense nutrients in the bulk, can enhance nutrient transport and drive interfacial instabilities \citep{atis2019microbial,narayanasamy2025metabolically}. Particle image velocimetry measurements and simulations of a hydrodynamic model show these metabolic flows manifest as a vortex ring in the bulk and a peak in the surface velocity outside of the colony. It remains, however, unclear how these flows, along with those generated by growth, connect to the colony morphology.

In the system considered in \cite{atis2019microbial} -- a microbial ``droplet'' growing on the surface of a viscous fluid -- growth generates tangential stresses at the surface and nutrient consumption drives bulk convection. The resulting fluid flows then feed back nonlocally onto the droplet dynamics. These features place the system within the broader class of problems involving rigid or deformable objects bound to a fluid interface. A foundational example is the translation of a rigid protein through a lipid membrane immersed in a viscous fluid, originally studied by Saffman and Delbr\"uck \citep{saffman1975brownian,saffman1976brownian} and later extended by many others \citep{hughes1981translational,stone1998hydrodynamics,stone2015mobility}. These works show that accounting for viscous stresses transmitted nonlocally through the bulk fluid is essential for predicting the particle mobility. The microbial system experiences analogous viscous stresses, generated not only by growth-induced motions, but also by solutal convection in the sublayer. Importantly, nonlocal hydrodynamic coupling fundamentally distinguishes this system from classical range expansion on rigid substrates in which the resistive force is proportional to the local cell velocity \citep{greenspan1976growth, lowengrub2009nonlinear, weady2024mechanics}.

A closer analog to the microbial system, in which the material itself deforms, is the free-boundary evolution of monolayer domains, such as surfactant-rich regions or lipid domains, which may relax under line tension, deform under electrostatic repulsion, or spread due to Marangoni stresses \citep{stone1995hydrodynamics,alexander2007domain,manikantan2020surfactant}. In more recent contexts, such motions may be driven by active processes within the monolayer domain itself, such as self-propulsion or rotation \citep{masoud2014collective,fei2017active,jia2022incompressible}. These additional degrees of freedom enable a wide range of instabilities and pattern-forming behaviors \citep{troian1990model,lubensky1996hydrodynamics,matar2009dynamics}. As in the Saffman-Delbr\"uck system, viscous forces imparted by the sublayer significantly modify the droplet mechanics, including the growth or decay rate of boundary perturbations and the form of internal stresses, which are generically singular at the boundary.

All of the systems above have a similar mathematical description. Typically, the bulk fluid velocity is described by a partial differential equation (PDE), such as the Stokes equations, coupled through boundary conditions to a constitutive law for the surface rheology, which may vary across domains. For domains with high degrees of symmetry, such as the disk or perturbations thereof, analytical solutions can often be found using integral transforms or eigenfunction expansions in body-fitted coordinates \citep{schneider1973slow,stone1995hydrodynamics}. When only integrated quantities, such as the total force or translational velocity, are desired, the Lorentz reciprocal theorem can bypass solution of the full flow field \citep{masoud2014reciprocal,stone2015mobility,elfring2016surface}. For non-structured geometries, a more flexible approach is to express the solution in terms of the associated Green's function of the bulk and/or surface operators \citep{lubensky1996hydrodynamics,chakrabarti2000solution,williams2004oblique,jia2022incompressible,askham2025integral}. Such formulations provide a generic representation of the droplet dynamics that can be simulated numerically. However, analytically determining the appropriate Green's function can be challenging. This difficulty is particularly acute in the microbial system, where metabolically generated buoyant flows induce a nonuniform body force in the bulk.

In this work, we develop and analyze a mathematical model for a microbial droplet growing, through consumption of an underlying nutrient, on the surface of a viscous fluid. Building on classical models of interfacial flows, the model takes the form of a free-boundary problem driven by three coupled components: (1) An isotropic pressure internal to the evolving droplet arising from growth, (2) bulk fluid flow driven by this growth pressure, and (3) buoyancy-driven flows arising from consumption of a nutrient. In section \ref{sec:model} we present the partial differential equations governing the full three-dimensional system. Dimensional analysis and asymptotic reduction simplifies these nonlinear equations to a set of linear PDEs. In section \ref{sec:BIE} we exploit this linearity to reformulate the model as a system of integro-differential equations defined solely on the droplet. In this formulation, several integral operators appear whose spectral properties we discuss in section \ref{sec:spectral}. These properties provide a clear route to axisymmetric solutions, which we derive in section \ref{sec:axisymmetric}. Finally, in section \ref{sec:stability} we analyze the geometric stability of these axisymmetric solutions. This analysis shows growth stabilizes axisymmetric geometries while flows induced by nutrient consumption amplify perturbations. These results offer a rigorous mechanical explanation of the morphological instabilities observed in experiments of growing yeast colonies \citep{atis2019microbial}, and provide a robust mathematical framework for the analysis of microbial communities growing on fluid interfaces. 

\section{A mathematical model}\label{sec:model}

\begin{figure}[t]
\centering
\includegraphics[scale=0.75]{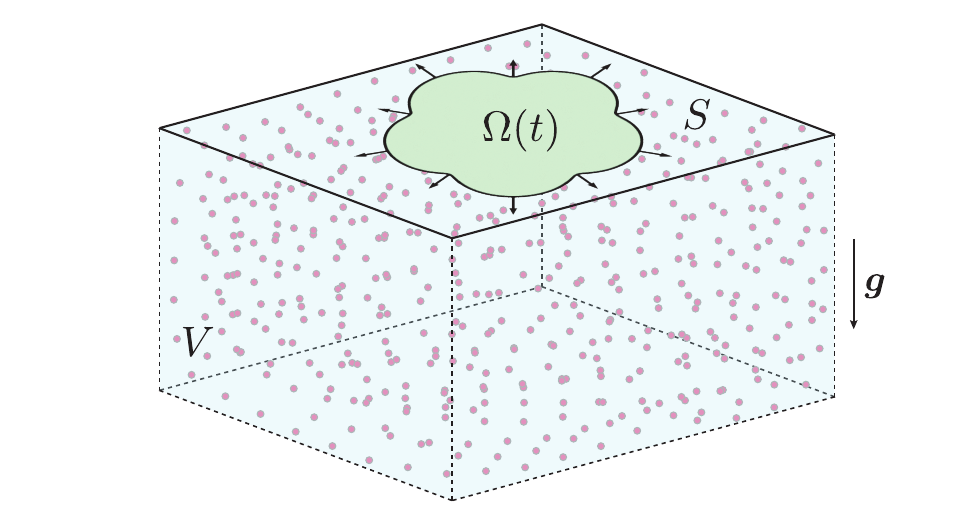}
    \caption{Schematic of the mathematical model. A growing microbial droplet $\Omega(t)$ (green) sits on the surface $S$ of a semi-infinite fluid volume $V$ (blue). As the droplet grows, it depletes dense nutrients (red circles) in the fluid which drives buoyant flows.}\label{fig:schematic}
\end{figure}

We consider a two-dimensional microbial droplet $\Omega(t) \subset S$ growing on the surface $S = \partial V = \{\bX = (x, y, 0) \in \bbR^3\}$ of a semi-infinite volume $V = \{\bX = (x, y, z) \in \bbR^3 : z < 0\}$ of a Newtonian fluid. A schematic of this configuration is shown in Fig. \ref{fig:schematic}. As the droplet expands, it exerts forces at the fluid surface which drive bulk flows. Simultaneously, microbial nutrient consumption establishes density gradients in the underlying fluid volume which lead to buoyancy-induced flows. The variables to be determined are therefore the bulk fluid velocity $\bU : V \to \bbR^3$ and pressure $P : V \to \bbR$, the nutrient concentration $c : V \to \bbR$ and fluid density $\rho(c) : V \to \bbR$, and the surface growth velocity $\bu : \Omega \to \bbR^2$ and growth pressure $p : \Omega \to \bbR$, which will be related through constitutive laws and boundary conditions. In the following we will need to restrict quantities in the fluid volume to the fluid surface. This is conveniently described by the operator $T$ which takes a vector field over $V$ to its tangential component on the surface, e.g. $T[(U_x(x,y,z), U_y(x, y, z), U_z(x, y, z))](x, y) = (U_x(x, y, 0), U_y(x, y, 0))$. 

\subsection{Fluid}

We assume the substrate is a Newtonian fluid with constant viscosity $\mu$ and variable density $\rho$. Conservation of momentum and mass in the fluid are
\begin{align}
    \rho\left(\frac{\partial\bU}{\partial t} + \bU\cdot\gradv\bU\right) = -\gradv P + \mu \lapv \bU - \rho g \be_z &\quad {\rm in} ~ V,\label{eq:dU/dt}\\
    \frac{\partial \rho}{\partial t} + \gradv\cdot(\rho \bU) = 0 &\quad {\rm in} ~ V\label{eq:divU},
\end{align}
where $g$ is acceleration due to gravity. Here $\gradv = (\partial_x, \partial_y, \partial_z)$ is the three-dimensional gradient and $\lapv = \partial_x^2 + \partial_y^2 + \partial_z^2$ is the three-dimensional Laplacian, with an analogous definition for $\grads$ and $\laps$.

The nutrient field $c$ within the fluid satisfies the advection-diffusion equation
\begin{equation}
    \frac{\partial c}{\partial t} + \gradv\cdot(c\bU) = D \lapv c \quad {\rm in} ~ V,\label{eq:dc/dt}
\end{equation}
where $D$ is a diffusion constant. For simplicity, we assume the density increases linearly with the nutrient field, $\rho(c) = \rho_0(1 + \alpha c)$, where $\rho_0$ is the fluid density in the absence of nutrients and $\alpha > 0$ is a constant. Making the Boussinesq approximation, Eqs. (\ref{eq:dU/dt})-(\ref{eq:divU}) become
\begin{align}
\rho_0\left(\frac{\partial\bU}{\partial t} + \bU\cdot\gradv\bU\right) = -\gradv P + \mu \lapv \bU - \rho_0 \alpha c g \be_z &\quad{\rm in} ~ V\label{eq:dU/dt-b},\\
\gradv\cdot\bU = 0 &\quad{\rm in} ~ V \label{eq:divU-b},
\end{align}
where we've absorbed the hydrostatic pressure: $P \mapsto P - \rho_0 g z$. 

\subsection{Growth}

Defining $\gamma(c)$ to be the microbial areal growth rate, the two-dimensional divergence of the surface velocity satisfies
\begin{equation}
    \grads \cdot \bu = \gamma(c) \quad {\rm in}~\Omega.\label{eq:gamma(c)}
\end{equation}
The functional form of $\gamma$ is a modeling choice but should be restricted by physical considerations. In particular, we expect $\gamma$ to be non-negative, non-decreasing, bounded, and satisfy $\gamma(0) = 0$. For the sake of completeness we do not yet specify an explicit form for $\gamma(c)$, however our analysis will be concerned with the nutrient-rich case $\gamma(c) = 1$. While this {\em a priori} violates the physical constraint $\gamma(0) = 0$, this choice is valid so long as the concentration is everywhere positive which will hold for sufficiently small domains.

\subsection{Boundary conditions}

On the droplet, growth forces are balanced by viscous forces. Letting $\bfv = [-P\bI_{\rm 3D} + \mu (\gradv\bU + \gradv\bU^T)]\cdot\be_z$ be the traction vector at the surface, we have
\begin{equation}
    \begin{cases}
        T[\bfv] = -\grads p & {\rm on} ~ \Omega,\\
        T[\bU] = \bu & {\rm on} ~ \Omega,\\
        \bU\cdot\be_z = 0 & {\rm on} ~ \Omega.
    \end{cases}\label{eq:ubc-Omega}
\end{equation}
The first condition describes the tangential force balance associated with the growth pressure $p$, the second is the no-slip condition, and the third requires that the surface remains flat. We assume line tension is negligible (see appendix \ref{app:line-tension}) so that the pressure vanishes along the boundary of the droplet, namely
\begin{equation}
    p = 0 \quad {\rm on} ~ \partial\Omega.\label{eq:pbc-Omega}
\end{equation}
Note that, because only $\grads p$ appears in the force balance (\ref{eq:ubc-Omega}), constant shifts of the pressure do not affect the solution. Moreover, the growth pressure is {\em not} the restriction of the three-dimensional fluid pressure $P$ to $S$. On the rest of the surface we have
\begin{equation}
    \begin{cases}
        T[\bfv] = \bzero & {\rm on} ~ S\setminus\overline\Omega,\\
        \bU\cdot\be_z = 0 & {\rm on} ~ S\setminus\overline\Omega,
    \end{cases}\label{eq:ubc-S}
\end{equation}
with suitable boundary conditions at infinity.

Outside of the droplet, the concentration field satisfies the boundary conditions
\begin{equation}
\begin{cases}
\frac{\partial c}{\partial z} = 0 & {\rm on} ~ S\setminus\overline\Omega,\\
c \to c_\infty & {\rm as} ~ |\bX| \to \infty.
\end{cases}
\end{equation}
The first equation says there is no nutrient flux on the open surface while the second sets nutrient supply in the far-field. On the droplet, the boundary condition arises from conservation of mass. Specifically, letting $\rho_b(\bx)$ be the microbial mass per unit area, which has units of ${\rm kg} / {\rm m}^2$, the flux of the total microbial mass $m_b = \int_\Omega \rho_b ~ \d\bx$ is
\begin{equation}
    \dot m_b = \int_\Omega \left(\frac{\partial \rho_b}{\partial t} + \grads\cdot(\bu\rho_b)\right) ~ \d\bx.
\end{equation}
Similarly, the flux of the nutrient mass $m_c = \int_V c ~ \d\bX$ is
\begin{equation}
    \dot m_c = D\left(\int_\Omega \gradv c\cdot\hat\bn ~ \d\bx\right).
\end{equation}
Enforcing $\dot m_b + \dot m_c = 0$, assuming $\rho_b$ is constant, and using $\grads\cdot\bu = \gamma(c)$, this yields the Robin-type boundary condition
\begin{equation}
D\frac{\partial c}{\partial z} = -\rho_b\gamma(c) \quad {\rm on}~\Omega.
\end{equation}

Finally, owing to growth, $\Omega$ will evolve in time. Letting $\bGamma(\theta) : [0, 2\pi) \to \partial \Omega$ be a parameterization of the boundary of $\Omega$, this evolution is given by the kinematic condition
\begin{equation}
    \frac{\partial\bGamma}{\partial t} = (\bu\rvert_{\bGamma}\cdot\hat\bn)\hat\bn,\label{eq:dgamma/dt}
\end{equation}
where $\hat\bn$ is the outward normal vector to $\Omega$, which is parallel to the surface $S$.

We note that this model is similar to that formulated in \cite{atis2019microbial}. In that model a no-slip condition $\bU = \bzero$ was applied on the droplet in place of the stress balance (\ref{eq:ubc-Omega}). In the axisymmetric case these boundary conditions should be equivalent when $\gamma \rightarrow 0$, though the coefficient $\rho_b/D$ must be scaled appropriately for the flux boundary condition to remain nontrivial.

\subsection{Dimensional analysis}\label{sec:dimensional-analysis}

The full set of equations is complex and nonlinear. However, through dimensional analysis and asymptotic approximations, we can reduce the system to a set of linear PDEs. Here we choose as a characteristic length scale $\ell_c = (|\Omega_0|/\pi)^{1/2}$ (e.g. the radius for an initially circular domain), time scale based on the characteristic growth rate $t_c = 1/\gamma_\infty$ with $\gamma_\infty = \gamma(c_\infty)$ the saturated growth rate, pressure scale $\mu / t_c$, and concentration scale $c_\infty$. Denoting dimensionless variables by primes, equations (\ref{eq:dU/dt-b})-(\ref{eq:divU-b}) in dimensionless form are
\begin{align}
    {\Rey}\left(\frac{\partial \bU'}{\partial t'} + \bU'\cdot\gradv'\bU'\right) = -\gradv' P' + \lapv'\bU' - {\Ray} c' \be_z & \quad {\rm in} ~ V',\\
    \gradv'\cdot\bU' = 0& \quad {\rm in} ~ V',
\end{align}
where ${\Rey} = (\ell_c^2 / t_c) / (\mu / \rho_0)$ is the Reynolds number, which compares the areal growth rate to momentum diffusion, and ${\Ray} = (\rho_0 \alpha c_\infty g \ell_c^3) / (\mu\ell_c^2/t_c)$ is the metabolic Rayleigh number, which compares buoyancy forces to growth forces. Similarly, the concentration equation (\ref{eq:dc/dt}) becomes
\begin{equation}
    {\Pec}\left(\frac{\partial c'}{\partial t'} + \bU'\cdot\gradv c'\right) = \lapv' c' \quad{\rm in} ~ V',
\end{equation}
where ${\Pec} = (\ell_c^2/t_c) / D$ is the P\'eclet number, which compares the areal growth rate to nutrient diffusion. The divergence condition on $\Omega$ becomes
\begin{equation}
\grads'\cdot\bu' = \gamma'(c') \quad{\rm on} ~ \Omega',
\end{equation}
and the corresponding flux condition is
\begin{equation}
\frac{\partial c'}{\partial z'} = -\beta \gamma'(c'),
\end{equation}
where $\beta = (\rho_b\gamma_\infty \ell_c) / (c_\infty D)$. (Note that here $\gamma'(c') = \gamma(c_\infty c')/\gamma_\infty$.) Finally, the concentration satisfies $c'\rightarrow 1$ in the far-field. All other boundary conditions keep the same form.

Assuming $\Rey, \Pec \ll 1$, the dimensionless velocity and pressure satisfy the forced Stokes equations
\begin{align}
    -\gradv' P' + \lapv' \bU' - {\Ray} c' \be_z = \bzero & \quad {\rm in} ~ V',\label{eq:stokes1}\\
    \gradv\cdot\bU' = 0 & \quad {\rm in} ~ V',\label{eq:stokes2}
\end{align}
and the dimensionless concentration satisfies the Laplace equation
\begin{equation}
    \lapv' c' = 0 \quad {\rm in} ~ V'.\label{eq:concentration}
\end{equation}
This asymptotic limit is significantly more tractable mathematically as the PDEs are linear and elliptic in the fluid volume, with all nonlinearity and time dependence coming from boundary conditions and geometry.

From a physical standpoint, the limit $\Rey, {\Pec} \ll 1$ holds when $\gamma_\infty|\Omega_0| \ll \mu/\rho_0, D$, which occurs for small microbial droplets with slow growth. We can estimate these parameters for the experiments of \cite{atis2019microbial} with the characteristic colony size $\ell_c = R \approx 2.5 ~ {\rm mm}$,  time scale $t_c = 1/\gamma_\infty \approx 1 ~ {\rm day}$, viscosity $\mu \approx 450 ~ {\rm Pa ~ s}$, density $\rho_0 \approx 10^3 ~ {\rm kg} / {\rm m}^3$ and solutal diffusivity $D \approx 2.4 \times 10^{-4} ~ {\rm mm}^2/{\rm s}$. This yields $\Rey \approx 1.6 \times 10^{-10}$ and $\Pec \approx 0.3$, both of which plausibly fall within the asymptotic regime. Moreover, using $\alpha c_\infty \approx 0.01$, the metabolic Rayleigh number at viscosity $\mu \approx 450 ~ {\rm Pa ~ s} $ is $\Ray \approx 50$. Finally, for the specific model $\partial c/\partial z = -c/\ell$ proposed in \cite{atis2019microbial} with depletion length $\ell \approx 2{\rm mm}$, we estimate $\beta \approx R/\ell \approx 1.3$. This will depend on the specific form of $\gamma$, but suggests $\beta = O(1)$ in general.

Experimentally, the viscosity ranges over about three orders of magnitude and strongly influences the interfacial dynamics. We will therefore ultimately be interested in the dependence of the solution on the metabolic Rayleigh number $\Ray$, which is inversely proportional to the viscosity. In the rest of this paper we assume all variables are dimensionless and we omit primes. 

\subsection{Comparison to classical interfacial flows}

This mathematical formulation is closely related to classical models of interfacial flows, such as those driven by Marangoni stresses in surfactant-laden interfaces \citep{manikantan2020surfactant} or by electrostatic effects in Langmuir films \citep{stone1995hydrodynamics}. In each of these cases, gradients in a surface field (e.g. the growth pressure, surface tension, or electrostatic potential) create stresses and drive fluid flow. As a consequence, the domain dynamics are inherently nonlocal. Despite these similarities, the present model has important differences. For one, the growth pressure arises from a differential constraint in a compactly supported region while other surface fields typically evolve through a surface PDE. Moreover, when that PDE is diffusive, such as in surfactant spreading at finite P\'eclet numbers, the field is distributed across the entire surface. Additionally, the surface and bulk are coupled not only through gradients in the growth pressure, but also through stresses induced by the buoyancy force $-\Ray c\be_z$, itself generated by surface-bulk coupling.

\section{Boundary integral formulation}\label{sec:BIE}

The three-dimensional system of equations above can, in principle, be studied analytically and/or solved numerically. However, the formulation couples PDEs defined on domains of different dimensions which significantly complicates analysis of the interfacial dynamics. Moreover, the formulation resists numerical solution as the bulk three-dimensional fluid must be discretized and coupled to a two-dimensional moving boundary problem on the surface. This is further complicated by the stress boundary condition on the open surface which generically results in singularities \citep{stephan1987boundary,costabel2003asymptotics}. To circumvent these issues, in this section we reformulate the problem as an integro-differential equation defined solely on the microbial droplet. We derive this equation by solving the system of equations (\ref{eq:stokes1})-(\ref{eq:concentration}) in Fourier space and computing the inverse transform of the solution evaluated on the surface. When buoyancy is neglected ($\Ray = 0$), we note that the Lorentz reciprocal theorem provides a simpler approach \citep{pozrikidis1992boundary}, however this cannot be exploited for general $\Ray$.

\subsection{Fourier space solution}\label{sec:BIE-Fourier}

Let $\bk \in \bbR^2$ be a wavevector with magnitude $k = |\bk|$. Taking the Fourier transform in the $xy$-plane of Eqs. (\ref{eq:stokes1})-(\ref{eq:concentration}), we find
\begin{align}
-\begin{pmatrix} \i\bk \\ \partial_z \end{pmatrix} \ft P + \left(-k^2 + \partial_z^2\right) \ft \bU - {\Ray} \ft c \be_z = \bzero,\\
\begin{pmatrix} \i\bk \\ \partial_z \end{pmatrix} \cdot \ft \bU = 0,\\
\left(-k^2 + \partial_z^2\right)\ft c = 0,
\end{align}
for $z < 0$, where
\begin{equation}
    (\ft\bU, \ft P, \ft c)(\bk, z) = \int_{\bbR^2} (\bU, P, c)(\bx,z) e^{-\i\bk\cdot \bx} ~ \d\bx
\end{equation}
denotes the Fourier transform with $\bx \in \bbR^2$. This linear system of ODEs can be solved using standard techniques, and the explicit solution can be found in appendix \ref{app:fourier}. Evaluating the solution at $z = 0$, we find a balance between the surface velocity, growth pressure, and concentration,
\begin{align}
\ft\bu_0 & = \i\bk\left( -\frac{\wft{\chi_\Omega p}}{2k} + \frac{\Ray}{8k^3} \ft c_0\right),\label{eq:u0-fourier}
\end{align}
where $\hat\bu_0(\bk) = \wft{T[\bU]}(\bk)$, $\hat c_0(\bk) = \hat c(\bk, 0)$ and $\wft{\chi_\Omega p} = \int_\Omega p e^{-\i\bk\cdot\bx}\d\bx$ with $\chi_\Omega$ the characteristic function, which naturally provides a zero extension
\begin{equation}
    (\chi_\Omega \nu)(\bx) = \begin{cases} \nu(\bx) & \bx \in \Omega, \\
    0 & \bx \in S\setminus\overline\Omega.
    \end{cases}
\end{equation}
Note that this calculation uses the boundary condition $p = 0$ on $\partial\Omega$, which allows us to commute $\wft{\chi_\Omega \grads p} = \i\bk\wft{\chi_\Omega p}$.

\subsection{An integro-differential equation}

Equation (\ref{eq:u0-fourier}) is expressed entirely in terms of quantities defined on the surface. However, the growth pressure is defined only on $\Omega$ while the concentration is defined on all of $S$. We can further represent $c$ by a function supported on $\Omega$ using the fundamental solution of the Laplace equation in 3D,
\begin{equation}
c(\bX) = 1 - \frac{1}{4\pi} \int_\Omega \frac{\sigma(\by)}{|\bX - \by|} ~ \d\by,\label{eq:c(X)}
\end{equation}
where $\sigma : \Omega \to \bbR$ is an {\it a priori} unknown density. This expression is harmonic (i.e. $\lapv c = 0$ in $V$), satisfies $\partial c / \partial z = 0$ on $S \setminus \Omega$, and obeys the far-field condition $c(\bX)\to 1$ as $|\bX|\to\infty$. The flux boundary condition $\partial c/\partial z = -\beta \gamma(c)$ on $\Omega$ can then be used to solve for $\sigma$. Specifically, noting that $\lim_{z\rightarrow 0^-} \partial_z \left( \int_\Omega \sigma(\by)/|(\bx,z)-(\by,0)| ~ \d\by\right) = 2\pi\sigma(\bx)$, we find an integral equation for $\sigma$, which may in general be nonlinear,
\begin{equation}
\sigma = 2\beta\gamma\left(1 - \cS_\Omega[\sigma]\right) \quad {\rm in} ~ \Omega,
\end{equation}
where
\begin{equation}
\cS_\Omega[\nu](\bx) = \frac{1}{4\pi}\int_\Omega \frac{\nu(\by)}{|\bx - \by|} ~ \d\by
\end{equation}
is the single layer potential of the three-dimensional Laplacian. Taking the Fourier transform of (\ref{eq:c(X)}) in the $xy$-plane for $\bk \neq \bzero$ gives $\ft c_0(\bk) = -\wft {\chi_\Omega\sigma}(\bk)/(2k)$ so that
\begin{equation}
\ft\bu_0  = -\i\bk\left(\frac{\wft{\chi_\Omega p}}{2k} + \frac{\Ray}{16k^4} \wft{\chi_\Omega\sigma}\right).\label{eq:u_0_hat}
\end{equation}
Taking the inverse Fourier transform and applying the convolution theorem (see appendix \ref{app:integral-operators}), we arrive at an integro-differential system of equations on $\Omega$:
\begin{align}
    \bu + \grads \left(\cS_\Omega[p] + \frac{\Ray}{16}\cB_{\Omega,\kappa}[\sigma] \right) = \bzero &\quad{\rm in} ~ \Omega,\label{eq:u}\\
    \grads\cdot\bu - \frac{\sigma}{2\beta} = 0 &\quad {\rm in} ~ \Omega,\label{eq:divu}\\
    \sigma - 2\beta\gamma\left(1 - \cS_\Omega[\sigma]\right) = 0 &\quad {\rm in} ~ \Omega,\label{eq:sigma}
\end{align}
where
\begin{align}
    \cB_{\Omega,\kappa}[\nu](\bx) &= \frac{1}{8\pi}\int_\Omega |\bx-\by|^2 \left[\log\left(\frac{|\bx-\by|}{R(\Omega)}\right) - 1 + \kappa\right] \nu(\by) ~\d\by
\end{align}
is the volume potential of the two-dimensional bilaplacian $\laps^2$. Here $R(\Omega) = (|\Omega|/\pi)^{1/2}$ is a characteristic length scale and $\kappa$ is an integration constant relating to an effective domain height; see appendix \ref{app:constant} for an explicit characterization. The system (\ref{eq:u})-(\ref{eq:sigma}) forms the complete set of integro-differential equations for the droplet velocity $\bu$, growth pressure $p$, and density $\sigma$.

In this system, the density $\sigma$ can be determined independently of $\bu$ and $p$ from Eq. (\ref{eq:sigma}). Moreover, taking the divergence of (\ref{eq:u}) and using (\ref{eq:divu}), we find
\begin{equation}
    \cN_\Omega[p] = -\left(\frac{\sigma}{2\beta} + \frac{\Ray}{16}\cV_{\Omega,\kappa}[\sigma]\right),\label{eq:N[p]}
\end{equation}
where the operators $\cN_\Omega$ and $\cV_{\Omega,\kappa}$ are
\begin{align}
    \cN_\Omega[\nu](\bx) &:= \laps \cS_\Omega[\nu](\bx) = \frac{1}{4\pi}\fpint_\Omega \frac{\nu(\by)}{|\bx - \by|^3} ~ \d\by,\\
    \cV_{\Omega,\kappa}[\nu](\bx) &:= \laps \cB_{\Omega,\kappa}[\nu](\bx) = \frac{1}{2\pi}\int_\Omega \log\left(\frac{|\bx - \by|}{R(\Omega)} + \kappa\right) \nu(\by) ~\d\by,
\end{align}
which are the hypersingular operator and the volume potential of the two-dimensional Laplacian, respectively. Here $\fpint$ denotes the Hadamard finite part integral, whose precise interpretation will be discussed in the following section. Given $\sigma$, Eq. (\ref{eq:N[p]}) can then be solved for $p$, independently of the velocity. Finally, once $p$ and $\sigma$ are determined, the velocity is evaluated using Eq. (\ref{eq:u}). For the following, we will suppress the $2{\rm D}$ subscript and use the notation $\nabla := \grads$ and $\Delta := \laps$. For concreteness we take $\kappa = 0$, which can be shown to effectively correspond to a depth $H \approx -1$ consistent with the experiments of \cite{atis2019microbial}; see appendix \ref{app:constant}.

Before proceeding, it is interesting to note that the integro-differential formulation closely resembles models of tumor growth and proliferating cell collectives on rigid substrates \citep{greenspan1976growth,lowengrub2009nonlinear, weady2024mechanics}. In the simplest case, these models take the form
\begin{align}
\bu + \nabla p = \bzero &\quad{\rm in} ~ \Omega,\label{eq:u_rigid}\\
\nabla\cdot\bu = \gamma &\quad{\rm in} ~ \Omega,\label{eq:divu_rigid}\\
p = 0 &\quad{\rm on} ~ \partial\Omega.\label{eq:pbc_rigid}
\end{align}
The present model differs critically in that the growth-induced pressure acts non-locally through the integral operator.

\section{Spectral analysis on the unit disk}\label{sec:spectral}

The integral operators $\cS_\Omega$ and $\cN_\Omega$ of the previous section commonly arise in boundary integral methods for open surfaces. When that surface is the unit disk $D = \{\bx\in\bbR^2 : |\bx| < 1\}$, their spectral properties are well characterized. In this section we review some of these properties which will be particularly useful in our analysis. We refer to \citep{wolfe1971eigenfunctions,boersma1993solution,martin1996mapping} for more thorough discussions. 

\subsection{Projected spherical harmonics}

To start, we introduce the (complex) projected spherical harmonics, which in polar coordinates $\bx(r,\theta) = r(\cos\theta, \sin\theta)^T$, $(r,\theta) \in [0,1]\times[0,2\pi)$, are given by
\begin{align}
    y^m_\ell(\bx) = \sqrt{ \frac{2\ell+1}{2\pi} \frac{(\ell - |m|)!}{(\ell + |m|)!} } P^{|m|}_\ell\left(\sqrt{1 - r^2}\right) e^{\i m \theta} , \quad \ell \geq 0 ~ {\rm and} ~ |m|\leq \ell,
\end{align}
where $P_\ell^m$ are the associated Legendre polynomials. Defining the weight function $\omega(\bx) = \sqrt{1 - |\bx|^2}$, when $\ell_1+m_1$ has the same parity as $\ell_2 + m_2$, these functions satisfy the orthogonality relation 
\begin{align}
\int_D \frac{y^{m_1*}_{\ell_1}(\bx) y^{m_2}_{\ell_2}(\bx)}{\omega(\bx)} ~ \d\bx =  \delta_{\ell_1, \ell_2}\delta_{m_1, m_2},
\end{align}
where $*$ denotes the complex conjugate. The subsets of functions with $\ell + m$ even/odd span two different spaces. Specifically, $\{y^m_\ell\}_{\ell + m ~ {\rm even}}$ is an orthonormal basis for $C^{\infty}(D)$ while  $\{y^m_\ell\}_{\ell+ m ~ {\rm odd}}$ is an orthonormal basis for the space $\{\nu(\bx) : \nu(\bx)/\omega(\bx) \in C^{\infty}(D)\}$.

\subsection{Generalized eigenfunctions and eigenvalues}

The functions $y_\ell^m$ with $\ell + m ~ {\rm even}$ and $\ell + m  ~ {\rm odd}$ are generalized eigenfunctions of the operators $\cS_D$ and $\cN_D$, respectively,
\begin{alignat}{2}
\cS_D\left[ \frac{y^m_\ell}{\omega} \right](\bx) &= \frac{\lambda^m_\ell}{4} \, y^m_\ell(\bx),  &&\quad \ell + m ~ {\rm even}, \label{eq:S_disk} \\
\cN_D\left[ y^m_\ell \right](\bx) &= -\frac{1}{\lambda^m_\ell} \, \frac{y^m_\ell(\bx)}{\omega(\bx)}, \label{eq:N_disk} &&\quad \ell + m ~ {\rm odd}, 
\end{alignat}
with generalized eigenvalues given by
\begin{equation}
\lambda_\ell^m \;=\; 
\frac{\Gamma\!\left(\tfrac{\ell+m}{2}+\tfrac{1}{2}\right)}
     {\Gamma\!\left(\tfrac{\ell+m}{2}+1\right)}
\frac{\Gamma\!\left(\tfrac{\ell-m}{2}+\tfrac{1}{2}\right)}
     {\Gamma\!\left(\tfrac{\ell-m}{2}+1\right)},
\end{equation}
where $\Gamma$ is the gamma function. Note that equations (\ref{eq:S_disk}) and (\ref{eq:N_disk}) imply $\cS_D$ maps function with $1/\omega(\bx)$ singularities to smooth functions, and $\cN_D$ maps functions that go to $0$ as $\omega(\bx)$ for $|\bx|\to 1$ to smooth functions. 

\subsection{Special cases}

In our analysis, we will often encounter the specific functions $y_m^m$ and $y_{m+1}^m$. For the former, using the fact that $P_m^m\left(\sqrt{1-r^2}\right) \propto r^m$, we find
\begin{equation}
\begin{aligned}
    y_m^m(\bx) \propto r^m e^{\i m\theta}.\label{eq:ymm}
\end{aligned}
\end{equation}
Next, using the recursion relation $P_{m+1}^m(x) = (2m+1)xP_m^m(x)$, we find
\begin{equation}
\begin{aligned}
    y_{m+1}^m(\bx) 
    & \propto \sqrt{1 - r^2} r^m e^{\i m\theta}.\label{eq:ymmp1}
\end{aligned}
\end{equation}
We will also need the products $(1 - r^2)r^m, (1 - r^2)^{3/2}r^m, (1 - r^2)^2r^m$ in terms of $P^m_{\ell}$ polynomials, which can be found in appendix \ref{app:Plm_expansion}. These relations also yield identities for the integral operators,
\begin{align}
    \cS_D[\omega(\zeta')\zeta'^j] &= \left(\frac{\lambda^j_j - \lambda^{j+1}_{j+1} |\zeta|^2 }{8}\right) \zeta^j, \label{eq:S[wz^j]}\\
    \cS_D[\omega^3(\zeta')\zeta'^j] &= \frac{3}{32} \left(
     \lambda^j_j - 2 \lambda^{j+1}_{j+1} |\zeta|^2 +  \lambda^{j+2}_{j+2} |\zeta|^4
    \right)\zeta^j,\label{eq:S[w^3z^j]}
\end{align}
and
\begin{align}
    \cN_D[\omega(\zeta') \zeta'^j] &= -\frac{1}{\lambda^j_{j+1}}\zeta^j,\label{eq:N[wz^j]}\\
    \cN_D[\omega^3(\zeta') \zeta'^j] &= \frac{3}{4\lambda^j_{j+1}} \left(-2 + \left(2 + \frac{1}{j+1}\right) |\zeta|^2\right) \zeta^j.\label{eq:N[w^3z^j]}
\end{align}
Finally, the eigenvalues satisfy the sum conditions
\begin{equation}
    \sum_{j=0}^m \lambda^j_j = (2m+1) \lambda^m_m,\label{eq:lambda_m^m}
\end{equation}
and
\begin{equation}
\sum_{j=0}^m \frac{1}{\lambda^j_{j+1}} = \frac{(2m + 3)}{3 \lambda^m_{m+1}},\label{eq:ilambda_m^m}
\end{equation}
which are straightforward to prove through induction.

\section{Axisymmetric solutions}\label{sec:axisymmetric}

Taking advantage of the spectral properties discussed in section \ref{sec:spectral}, we now seek axisymmetric solutions to the system of integro-differential equations (\ref{eq:u})-(\ref{eq:sigma}) on the disk of radius $R$, denoted by $D_R = \{ \bx \in \bbR^2 : |\bx| < R\}$. For simplicity, we assume we are in the nutrient-rich regime and take $\gamma(c) = 1$. We individually solve the problems for growth-dominated flow (${\Ray} = 0$) and buoyancy-dominated flow (${\Ray} \to \infty$) from which we can construct the solution for general $\Ray$.

\subsection{Growth}

We first consider the case of growth-dominated flow, ${\Ray} = 0$, for which the system of equations reduces to
\begin{align}
    \bug + \nabla \cS_{D_R}[\pg] &= \bzero,\label{eq:u0_g}\\
    \cN_{D_R}[\pg] + 1 &= 0.\label{eq:p0_g}
\end{align}
Making the change of variables $\bx = R\bx'$, $\bug = R\bug'$ and $\pg = R\pg'$, the equations transform to
\begin{align}
    \bug' + \nabla' \cS_D[\pg'] &= \bzero,\label{eq:u0_g_unit}\\
    \cN_{D}[\pg'] + 1 &= 0.\label{eq:p0_g_unit}
\end{align}
These equations are independent of $R$, implying the solution for $(\bug, \pg)$ is homogeneous in $R$. Noting that $y_1^0 \propto \omega$ and using the property $\cN_D[y_\ell^m] = -y_\ell^m/(\lambda_\ell^m \omega)$, the scaled pressure is given by
\begin{align}
    \pg'(\bx') = \lambda_1^0\sqrt{1 - |\bx'|^2}.
\end{align}
We can in principle evaluate $\cS_D[\pg']$ to compute $\bug'$. However, under the assumption of axisymmetry, the velocity can be derived directly from the condition $\nabla'\cdot\bug' = 1$ so that
\begin{equation}
    \bug'(\bx') = \frac{\bx'}{2}.
\end{equation}

\subsection{Buoyancy}

We next consider the limit of buoyancy-dominated flow $\Ray \to \infty$. Defining $\bu = (\beta{\Ray}) \bub$, $p = (\beta{\Ray}) \pb$, and $\sigma = \beta\sigma_b$ and taking $\Ray\to\infty$, the system of equations is
\begin{align}
    \bub + \nabla\cS_{D_R}[\pb] + \frac{1}{16}\nabla\cB_{D_R}[\sigmab] &= \bzero,\\
    \cN_{D_R}[\pb] + \frac{1}{16} \cV_{D_R}[\sigmab] &= 0,\\
    \sigma_b - 2 &= 0.
\end{align}
Again defining $\bx = R\bx'$, $\bub = R^3\bub'$, $\pb = R^3\pb'$, and $\sigmab = \sigmab'$, we get
\begin{align}
    \bub' + \nabla'\cS_D[\pb'] + \frac{1}{16}\nabla'\cB_D[\sigmab'] &= \bzero,\label{eq:u0_b}\\
    \cN_D[\pb'] + \frac{1}{16}\cV_D[\sigmab'] &= 0,\label{eq:p0_b}\\
    \sigma_b' - 2 &= 0,\label{eq:s0_b}
\end{align}
which is independent of $R$.

\begin{figure}[t]
    \centering
    \includegraphics[width=\linewidth]{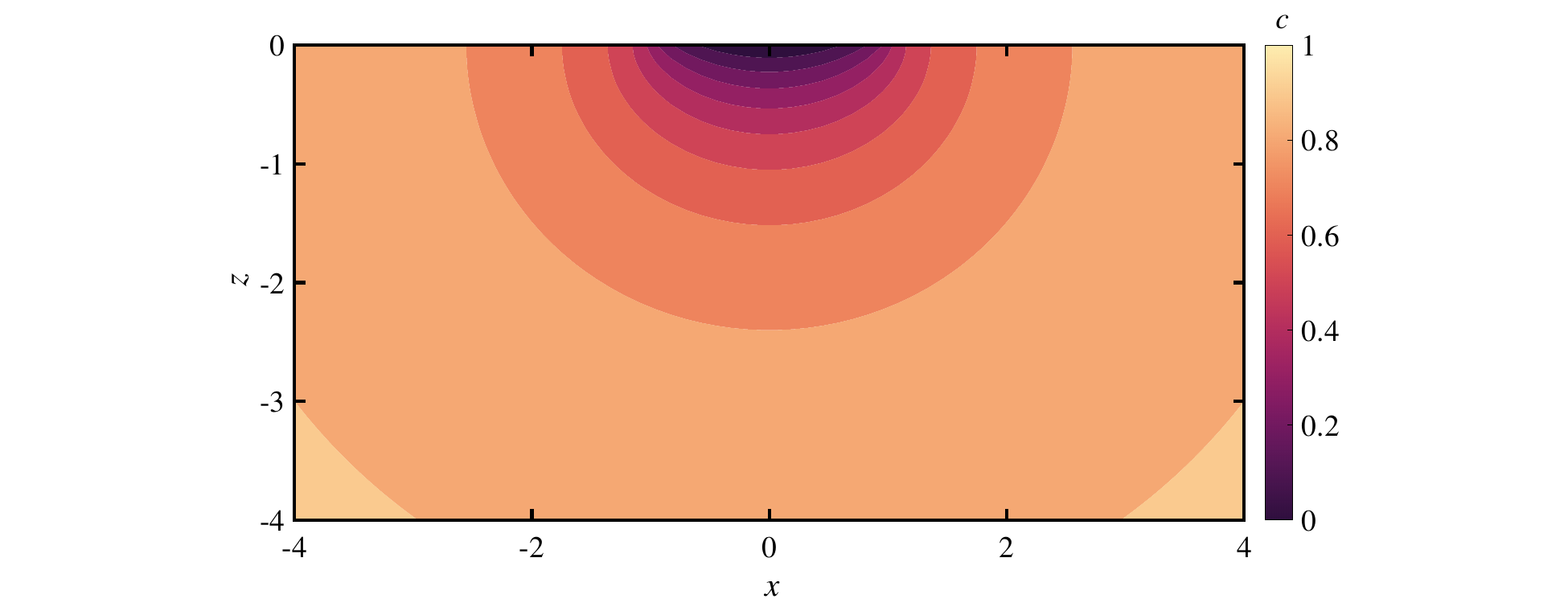}
    \caption{Axisymmetric solution for the nutrient concentration field. Nutrients are depleted near the droplet and the concentration increases in concentric ellipsoidal-like contours which become increasingly circular.}

    \label{fig:concentration}
\end{figure}

Now trivially we have $\sigma_b'(\bx') = 2$, which we can use to numerically evaluate the concentration field in the bulk $V$. A cross-sectional view of the corresponding concentration field is shown in Fig. \ref{fig:concentration}, where we see nutrient is depleted in the vicinity of the droplet and increases outward in the form of concentric ellipsoidal-like contours which become increasingly circular at far distances.

To compute the pressure, we need to evaluate $\cV_D[\sigmab']$. Using Eq. (\ref{eq:V_D[z^m]}), we find
\begin{align}
    \cV_D[\sigmab'](\bx') &= \frac{|\bx'|^2 - 1}{2}.\label{eq:V_D[sigma]}
\end{align}
The pressure therefore solves the equation
\begin{equation}
    \cN_D[\pb'](\bx') = -\left(\frac{|\bx'|^2 - 1}{32}\right).
\end{equation}
Using the identities (\ref{eq:N[wz^j]}) and (\ref{eq:N[w^3z^j]}), we find
\begin{equation}
\pb'(\bx') = -\frac{\lambda_1^0\sqrt{1 - |\bx'|^2}}{96}\left(1 + \frac{4}{3}(1 - |\bx'|^2)\right).\label{eq:pb}
\end{equation}
Finally, as before, the axisymmetric velocity can be determined directly from the condition $\nabla'\cdot\bub' = 0$, from which we immediately see $\bub' = \bzero$ in $D$. 

\begin{figure}[t]
    \centering
    \includegraphics[width=\linewidth]{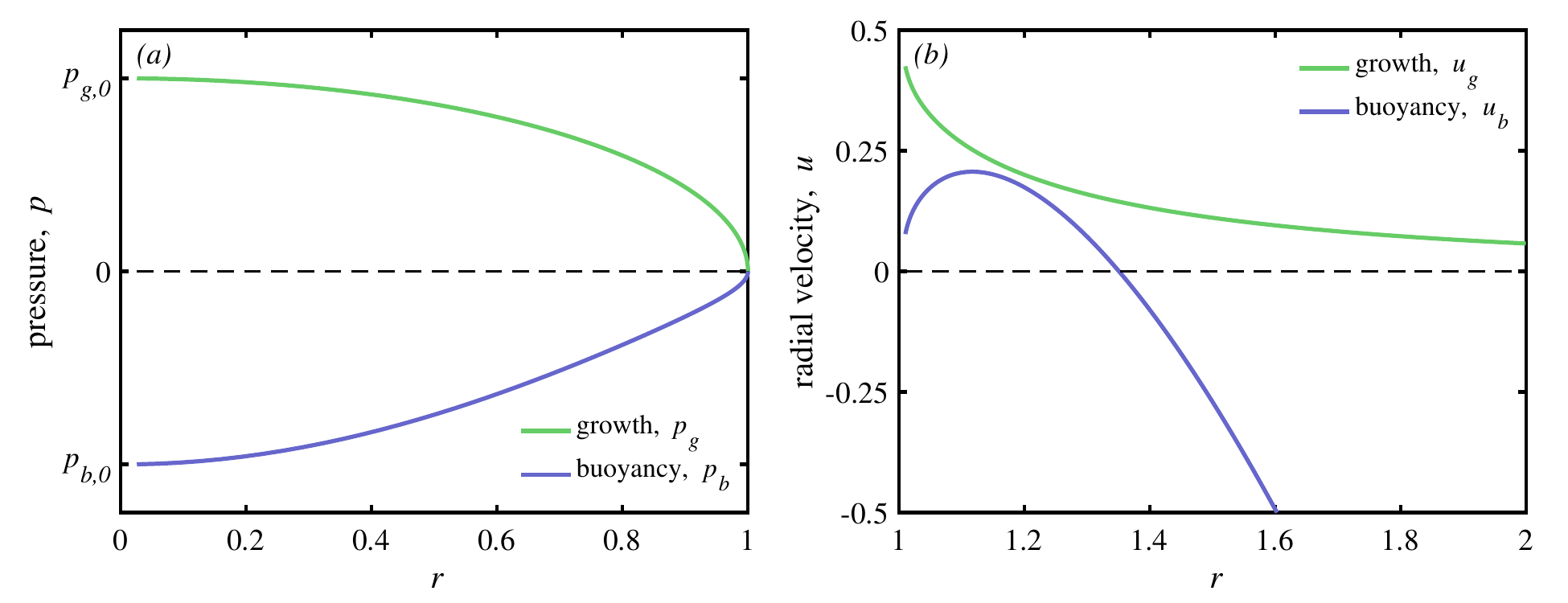}
    \caption{{\it (a)} Axisymmetric pressure on the unit disk for growth-dominated flow (green) and buoyancy-dominated flow (purple), each scaled by their value at the origin ($p_{g,0} \approx 1.27$ and $p_{b,0} \approx -0.031$) for visualization. The former is strictly positive while the latter is strictly negative. {\it (b)} Radial velocity for growth-dominated flow (green) and buoyancy-dominated flow (purple) exterior to the droplet. The former is strictly positive and monotonically decreasing while the latter exhibits a local maximum near $r \approx 1.1$.}
    \label{fig:axisymmetric-surface}
\end{figure}

\subsection{General axisymmetric solution}\label{ss:general-axisymmetric}

For pure growth (${\Ray} = 0$) we found $\bug = R\bug'$ and $\pg = R\pg'$, where $(\bug',\pg')$ are the solution on the unit disk. For buoyancy-dominated flow $({\Ray} \to \infty)$, we found $\bub = R^3\bub'$ and $\pb = R^3\pb'$ where $(\bub', \pb')$ are the solution on the unit disk with unit $\Ray$, unit $\beta$, and a divergence-free velocity. Using linearity of Eqs. (\ref{eq:u})-(\ref{eq:sigma}), the general solution for the disk of radius $R$ may then be written as
\begin{equation}
    p(\bx) = R\pg'(\bx/R) + R^3\beta{\Ray}\pb'(\bx/R).
\end{equation}
along with the velocity
\begin{equation}
    \bu(\bx) = \frac{\bx}{2}.\label{eq:u-axi}
\end{equation}
From this expression for the velocity along with the kinematic boundary condition (\ref{eq:dgamma/dt}), this shows this axisymmetric solution is a radially expanding disk whose radius grows exponentially in time, $R(t) = R(0)e^{t/2}$.

Figure \ref{fig:axisymmetric-surface}{\it (a)} shows the axisymmetric pressure due to growth $p_g$ and buoyancy $p_b$ on the unit disk with $\beta = 1$, scaled by the magnitude of their values at the origin for visualization ($p_{g,0} \approx 1.27$ and $p_{b,0} \approx -0.031$). The former is strictly positive while the latter is strictly negative. Thus, as the Rayleigh number increases, the pressure will eventually take on negative values. Moreover, the pressure becomes negative everywhere inside $D$ when $\beta\Ray \geq 96$.

The radial velocity on the surface outside of the disk for both growth- and buoyancy-dominated flow, evaluated numerically, is shown in Fig. \ref{fig:axisymmetric-surface}{\it (b)}. For visualization, the buoyancy-dominated flow is scaled by a factor of 96. For growth-dominated flow, the velocity decreases monotonically from $u_g = 1/2$ at $r = 1$, consistent with continuity of the velocity, towards $u_g \rightarrow 0$  as $r\rightarrow\infty$. The scaling here is $O(1 / r^2)$ owing to the $1/r$ kernel in $\cS_D$. For buoyancy-dominated flow, the velocity initially increases from $u_b = 0$ at $r = 1$, again consistent with continuity, and later decreases after reaching a maximum at $r \approx 1.1$. This maximum is also observed in experiments \citep{atis2019microbial}. The velocity in this case eventually takes on negative values and scales as $u = O(r\log r)$ owing to the kernel in the operator $\cB_D$. This nonphysical divergence of the velocity is similar to the Stokes paradox and arises from the zero $\Rey$ and $\Pec$ approximation on an unbounded domain. Hence the solution should be interpreted as an inner solution which is matched to an outer solution by the free constant $\kappa$ in the operators $\cB_{\Omega,\kappa}$ and $\cV_{\Omega,\kappa}$.

\begin{figure}[t]
    \centering
    \includegraphics[width=\linewidth]{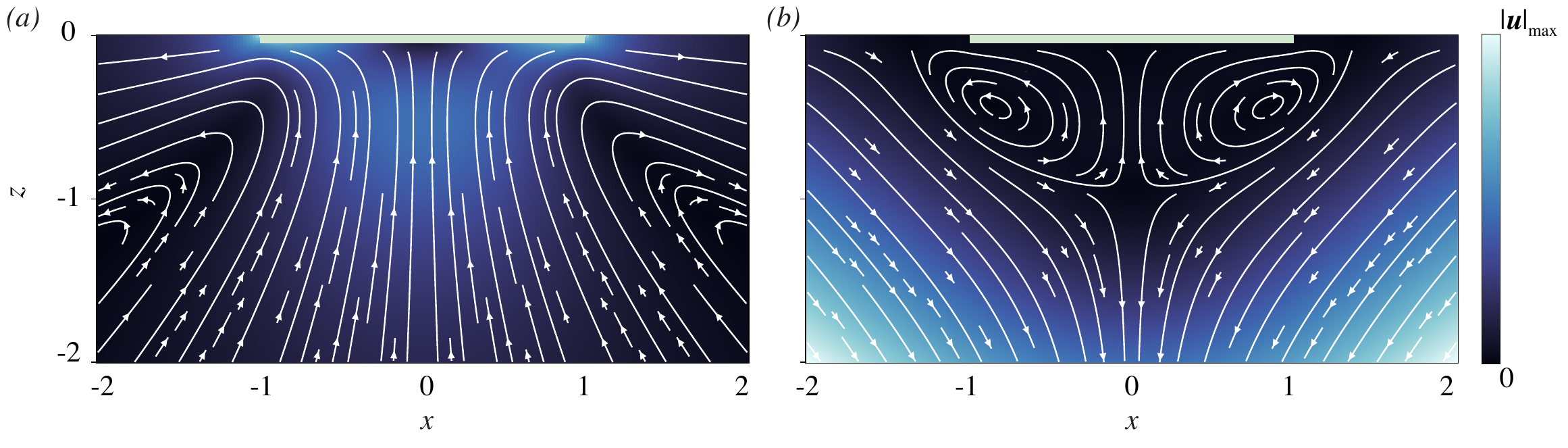}
    \caption{Cross-sectional streamlines and velocity magnitude in the {\it (a)} growth-dominated regime ($\Ray = 0$) and {\it (b)} buoyancy-dominated regime ($\Ray\rightarrow\infty$) for $\kappa = 0$. In the former, fluid is pulled from below and pushed outward, reminiscent of stagnation point flow. In the latter, a vortex ring forms below the droplet, with the core positioned near the droplet edge and a stagnation point at $z \approx -1$.} \label{fig:axisymmetric-bulk}
\end{figure}

A cross sectional view of the streamlines and velocity field in the bulk is shown in Fig. \ref{fig:axisymmetric-bulk} for {\it (a)} growth-dominated flow and {\it (b)} buoyancy-dominated flow. The formulation used for these computations is described in appendix \ref{app:integral-operators}. The flow field in the growth-dominated regime resembles a stagnation point flow, where fluid is pulled from beneath the droplet and pushed outward at the surface. The character of the flow changes significantly in the buoyancy-dominated regime, where a vortex ring forms beneath the droplet. This vortex ring extends slightly beyond the droplet, with its core located near the droplet boundary. Note that the extent of this vortex ring is modified by the constant $\kappa$, which sets the position of the lower stagnation point.

\section{Linear stability analysis}\label{sec:stability}

In this section, we analyze the geometric stability of the axisymmetric solution in the nutrient-rich regime $\gamma(c) = 1$ at short times. To assess the evolution of small perturbations to this solution, we make an Ansatz for a parameterization of the boundary $\partial\Omega_\eps^m(t)$ of the perturbed domain $\Omega_\eps^m(t)$,
\begin{equation}
\bGamma^m(\theta, t) = R(t)\left[\left( 1 + \eps(t) \cos(m\theta) \right)\hat\br + \eps(t) \sin (m\theta) \hat\btheta\right], \quad {\rm for} ~ \theta \in [0, 2\pi),\label{eq:gamma_eps^m}
\end{equation}
where $\eps(t) \ll 1$ is the perturbation amplitude, $m$ is the perturbation frequency, and $\hat\br = (\cos\theta, \sin\theta)^T$ and $\hat\btheta = (-\sin\theta, \cos\theta)^T$ are basis vectors. We will show that, under a suitable tangential velocity which does not affect the geometry, this Ansatz is preserved up to $O(\eps^2)$. As in the axisymmetric solution, we separate our analysis into the growth-dominated (${\Ray} = 0$) and buoyancy-dominated (${\Ray} \to\infty$) regimes from which we can reconstruct the solution for general ${\Ray}$.

\subsection{Representation on the unit disk}

Because we are considering geometric deformations and the integral operators depend on the domain, it is critical to define the equations over a fixed reference domain, which we take to be the unit disk $D = \{\bxi \in \bbR^2 : |\bxi| < 1\}$. To this end, let $\bfv : D \to \Omega$ be an invertible mapping and let $\tilde\be_i(\bxi) = \partial\bfv/\partial\xi_i$ be the associated (unnormalized) basis vectors for $i = 1,2$, where $\bxi = (\xi_1, \xi_2) \in D$. This yields the metric tensor $(\bG_{\bfv})_{ij} = \tilde\be_i\cdot\tilde \be_j$ and Jacobian $J_{\bfv}(\bxi) = \sqrt{\det\bG_{\bfv}(\bxi)}$. The system of equations is then
\begin{align}
    \tbu + \bG_{\bfv}^{-1}\left(\tnabla \cS_{\bfv}[p] + \frac{\Ray}{16} \tnabla  \cB_{\bfv}[\sigma]\right) = \bzero &\quad{\rm in} ~ D,\\
    J_{\bfv}^{-1}\tnabla \cdot(J_{\bfv}\tbu) - \frac{\sigma}{2\beta} = 0 &\quad{\rm in} ~ D,\\
    \sigma - 2\beta = 0 &\quad{\rm in} ~ D,
\end{align}
where $\tilde\bu = (\tilde u_1,\tilde u_2)$ is expressed in the basis $\tilde\be_1,\tilde\be_2$ such that $\bu = \tilde u_1\tilde\be_1 + \tilde u_2 \tilde\be_2$ and $\tnabla$ denotes the gradient operator on $D$. The transformed integral operators are
\begin{align}
    \cS_{\bfv}[\nu](\bxi) &= \frac{1}{4\pi}\int_D \frac{1}{|\bfv(\bxi) - \bfv(\bxi^\prime)|   } \left(\nu \circ \bfv\right)(\bxi^\prime)J_{\bfv}(\bxi^\prime) ~ \d\bxi^\prime, \\
    \cB_{\bfv}[\nu](\bxi) &= \frac{1}{8\pi}\int_D |\bfv(\bxi) - \bfv(\bxi^\prime)|^2\left[\log\left(\frac{\left|\bfv(\bxi) - \bfv(\bxi^\prime)\right|}{R(\Omega)}\right) - 1\right]\left(\nu \circ \bfv\right)(\bxi^\prime)J_{\bfv}(\bxi^\prime) ~ \d\bxi^\prime.
\end{align}
Similarly, the pressure satisfies
\begin{align}
    \cN_{\bfv}[p] = -\left(\frac{\sigma}{2\beta} + \frac{\Ray}{16} \cV_{\bfv}[\sigma]\right),
\end{align}
where
\begin{align}
    \cN_{\bfv}[\nu](\bxi) &= \frac{1}{4\pi}\fpint_D \frac{1}{|\bfv(\bxi) - \bfv(\bxi^\prime)|^3} \left(\nu \circ \bfv\right)(\bxi^\prime)J_{\bfv}(\bxi^\prime) ~ \d\bxi^\prime, \\
    \cV_{\bfv}[\nu](\bxi) &= \frac{1}{2\pi} \int_D \log\left(\frac{\left|\bfv(\bxi) - \bfv(\bxi^\prime)\right|}{R(\Omega)}\right) \left(\nu \circ \bfv\right)(\bxi^\prime)J_{\bfv}(\bxi^\prime) ~ \d\bxi^\prime.
\end{align}

While in principle $\bfv$ can be any invertible map, it is particularly advantageous to choose it to be conformal so that $\bG_{\bfv}(\bxi) = J_{\bfv}(\bxi)\bI$. This choice allows us to introduce the complex variable $\zeta = \xi_1 + \i\xi_2$ associated to the point $\bxi = (\xi_1, \xi_2) \in D$ and the associated complex-valued conformal map $f : D \to \Omega$. Moreover, it preserves the form of the singularities in the kernel, allowing us to compute shape derivatives \citep{martin1996mapping}. Geometrically, the conformal map also preserves angles so that the normal component of $\bu$ is preserved, i.e.
\begin{equation}
    (\bu\cdot\hat\bn)(f(e^{\i \theta})) = (J_f^{1/2}\tbu\cdot\hat\br)(e^{\i \theta}).
\end{equation}

It is readily seen that
\begin{equation}
\feps(\zeta, t) = R(t)[\zeta + \eps(t) \zeta^{m+1}]
\end{equation}
is a conformal map from $D$ to $\Omega_\eps^m(t)$. The associated basis functions may then be expressed in the polar coordinate system $\zeta = re^{\i \theta}$ as
\begin{align}
    \tilde\be_r &= 
    R\left[(1 + \eps(m + 1) r^m \cos m\theta)\hat\br + \eps(m + 1) r^m \sin m\theta \hat\btheta \right],\\
    \tilde\be_\theta &= 
    R\left[ -\eps(m+1)r^{m+1}\sin m\theta \hat\br + (1 + \eps(m + 1)r^{m+1}\cos m\theta)\hat\btheta\right],
\end{align}
which notably scale in proportion to $R$. The normal vector, in terms of $\theta$, is then
\begin{equation}
    \hat\bn = \frac{\tilde\be_r}{|\tilde\be_r|}\Big\rvert_{r = 1} = \hat\br + \eps(m + 1)\sin m \theta \hat\btheta + O(\eps^2).
\end{equation}
We similarly define the scaling transformation $f_R(\zeta, t) = R(t)\zeta$, which describes the radially expanding base state. 

\subsection{Linearization}

In the following analysis we will need the shape derivative, with respect to $\eps$, of the Jacobian and the integral operators induced by the conformal map. We will use the natural complexification for which complex-valued functions are interpreted as their real part. The Jacobian is straightforward,
\begin{align}
    \Jeps{R} = \lim_{\eps \to 0}\frac{J_{\feps}-J_{f_R}}{\eps} =  2(m+1)R^2 \zeta^m.
\end{align}
For an integral operator $\cT_{\feps}$, we adopt the following notation for the shape derivative with respect to $\eps$ at fixed $R$,
\begin{align}
    \cTeps{R}[\nu] = \lim_{\eps \to 0 } \frac{\cT_{ \feps}\left[\nu\circ\left(\feps\right)^{-1}\right] - \cT_{f_R}\left[\nu \circ f_R^{-1}\right]}{\eps}.
\end{align}
Using the identity
\begin{align}
    \frac{ \feps(\zeta) - \feps(\pzeta)}{\zeta - \pzeta}  = R\left(1 + \eps \sum_{j = 0}^m \zeta^j \pzeta^{m-j}\right) + O(\eps^2),
\end{align}
a straightforward, but technical, calculation gives
\begin{align}
    \cSeps{R}[\nu](\zeta) &= R
    \left[ 2(m+1) \cS_D[ \pzeta^m \nu(\pzeta)] -  \cS_D[C_m(\zeta, \pzeta) \nu(\pzeta)] \right], 
     \\
    \cNeps{R}[\nu](\zeta) &=  R^{-1}\left[ 2(m+1) \cN_D[ \pzeta^m \nu(\pzeta)] - 3 \cN_D[C_m(\zeta,\pzeta) \nu(\pzeta)](\zeta) \right], \\
    \cVeps{R}[\nu](\zeta) &=
    R^2\left[2(m + 1)\cV_D[\zeta'^m \nu(\zeta')] + \cI_{D,0}[C_m(\zeta, \zeta') \nu(\zeta')]\right],\\
    \cBeps{R}[\nu](\zeta) &=
    R^4\Big[ 2(m + 1)\cB_D[\zeta'^m\nu(\zeta')] + 2\cB_D[C_m(\zeta, \zeta')\nu(\zeta')] 
    \nonumber 
    \\ &  \hspace{6ex} + (1/4)\cI_{D,2}[C_m(\zeta,\zeta')\nu(\zeta')]\Big],
    \end{align}
where
\begin{equation}
    C_m(\zeta, \pzeta) = \sum_{j=0}^m \zeta^j \pzeta^{m-j}
\end{equation}
and
\begin{equation}
    \cI_{D,k}[\nu] = \frac{1}{2\pi}\int_D |\zeta - \zeta'|^k \nu(\zeta') ~ \d\zeta'.
\end{equation}
For a function of two variables $\nu(\zeta, \zeta')$ and integral operator $\cT_D$ with kernel $K(\zeta - \zeta')$, we use the slight abuse of notation
\begin{equation}
    \cT_D[\nu(\zeta, \zeta')](\zeta) = \int_D K(\zeta - \zeta')\nu(\zeta, \zeta') \d\zeta',
\end{equation}
where primed variables always denote the integration variable.

\subsection{Growth}

We first analyze the stability of growth-dominated flow, ${\Ray} = 0$. Here the system of equations on the reference domain is
\begin{align}
    \tbu_g + J^{-1}_{\feps}\tilde\nabla \cS_{\feps}[p_g] &= \bzero,\\
   \cN_{\feps}[p_g] + 1 &= 0.
\end{align}
We assume a regular perturbation expansion for each variable,
\begin{align}
\tbu_g &= \tbu_0 + \eps \tbu_1 + O(\eps^2), \\
(p_g\circ\feps) &= R(p_0 + \eps p_1) + O(\eps^2).
\end{align}
(Recall $\tbu$ is expressed in the unnormalized basis so that $\bu = \tilde u_1 \tilde \be_1 + \tilde u_2 \tilde\be_2$ scales with $R$.) Substituting above and matching powers of $\eps$, the $O(1)$ equations are given by the unit disk equations
\begin{align}
    \tbu_0 + \tnabla\cS_D[p_0] &= \bzero,\\
    \cN_D[p_0] + 1 &= 0,
\end{align}
and the $O(\eps)$ equations are
\begin{align}
    \tbu_1 + \Jeps{} \tbu_0 + \tnabla \cS_D[p_1] + \tnabla \cSeps{}[p_0] &= \bzero, \label{eq:ueps_g}\\
    \cN_D[p_1] + \cNeps{}[p_0] &= 0,\label{eq:peps_g}
\end{align}
where we used $\tbu_0 + \tnabla \cS_D[p_0] = \bzero$ from the $O(1)$ equation. 

To compute the $O(\eps)$ pressure $p_1$, we will need to evaluate $\cNeps{}[p_0]$. In section \ref{sec:axisymmetric}, we showed $p_0(\zeta) = \lambda_1^0\omega(\zeta)$. Using the identities (\ref{eq:N[wz^j]}) and (\ref{eq:ilambda_m^m}), we find
\begin{equation}
\cNeps{}[p_0] = -\frac{\lambda_1^0}{\lambda_{m+1}^m}\zeta^m.
\end{equation}
From this, using the fact that $y_{m+1}^m(\zeta) \propto  \omega(\zeta)\zeta^m$, we can solve Eq. (\ref{eq:peps_g}) for $p_1$,
\begin{align}
     p_1(\zeta) = \lambda^0_1 \omega(\zeta){\zeta^m}.
\end{align}
To determine the velocity, we need to evaluate $\cS_D[p_1]$ and $\cSeps{}[p_0]$. The identity (\ref{eq:S[wz^j]}) directly gives
\begin{equation}
    \cS_D[p_1] = \lambda^0_1\left(\frac{\lambda^{m}_{m} - \lambda^{m+1}_{m+1} |\zeta|^2}{8}\right) \zeta^m.
\end{equation}
Finally, using (\ref{eq:S[wz^j]}) and (\ref{eq:lambda_m^m}) we find
\begin{equation} 
\cSeps{}[p_0] = \lambda^0_1\left(\frac{\lambda^{m}_{m} - \left(\lambda^0_0 - \lambda^{m+1}_{m+1}\right) |\zeta|^2}{8}\right)\zeta^m.
\end{equation}
This yields the $O(\eps)$ interface velocity
\begin{equation}
\begin{aligned}
   (\tbu_1 \cdot \hat{\br})(e^{\i \theta}) &= -\Real \left[(m+1) \zeta^m (\tbu_0\cdot\hat{\br}) + \frac{\partial}{\partial r}\left(\cS_D[p_1] + \cSeps{}[p_0]\right) \right ]\Big\rvert_{\zeta = e^{\i \theta}}\\
   &= \left(\frac{1}{2} - m\frac{\lambda_1^0\lambda_m^m}{4}\right) \cos m\theta
   \\ & =: \sigma_g^m \cos m\theta.
\end{aligned}
\end{equation}

\begin{figure}[t]
    \centering
    \includegraphics[width=\linewidth]{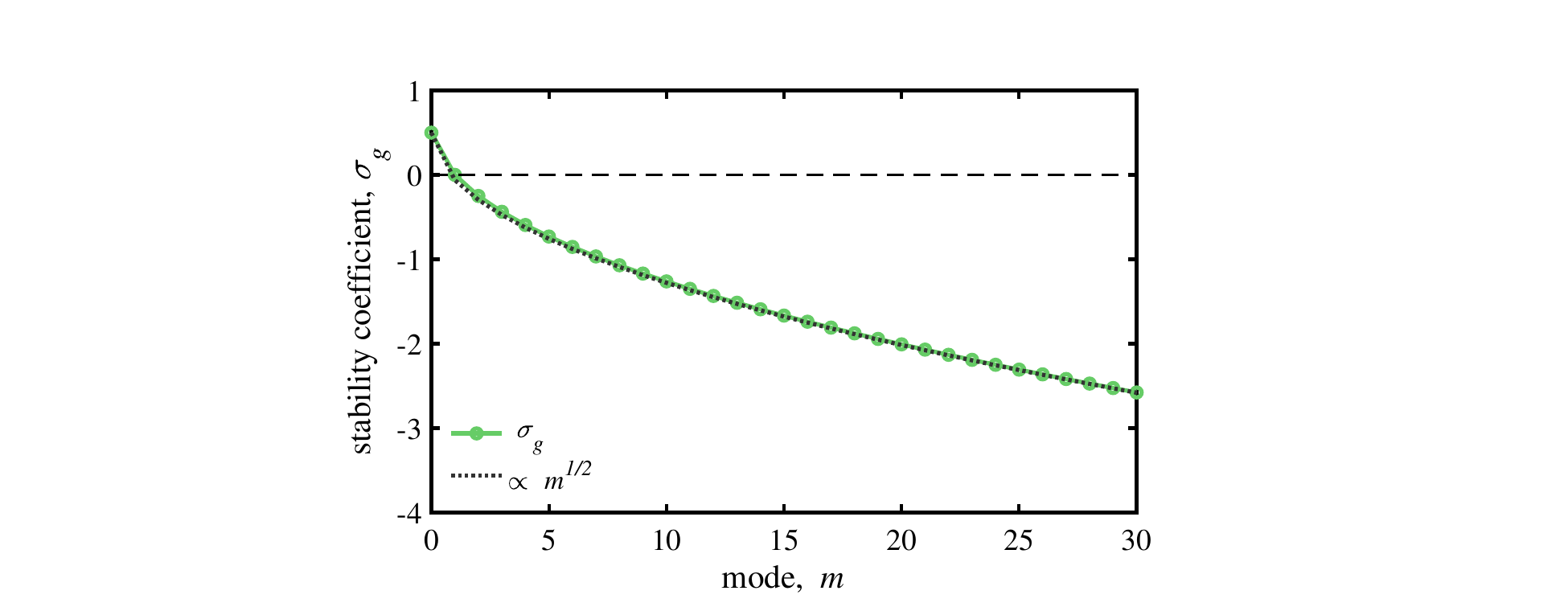}
    \caption{Stability coefficient $\sigma_g^m$ for growth-dominated flow ($\Ray = 0$). For $m = 0$, which corresponds to a dilation, the coefficient is $\sigma_g^0 = 1/2$, consistent with the $O(1)$ solution $R(t) = R_0e^{t/2}$. For $m = 1$ we have $\sigma_g^1 = 0$, which is reflective of translational invariance. All other coefficients are negative, indicating that growth is stabilizing, and tend towards negative infinity as $O(m^{1/2})$.}
    \label{fig:sigma_g}
\end{figure}

Values of the coefficient $\sigma_g^m$ greater than $1/2$ indicate that perturbations are amplified while negative values indicate they are suppressed. (Note that $\sigma_g^m$ is not strictly the growth rate associated with the perturbation owing to the growing domain; see Eq. (\ref{eq:eps(t)}) for an explicit characterization.) A plot of $\sigma_g^m$ is shown in Fig. $\ref{fig:sigma_g}$. The $m = 0$ mode has $\sigma_g^m = 1/2$ and reflects a dilation. The $m = 1$ mode has $\sigma_g^m = 0$, which is consistent with translational invariance. All other modes are negative, showing growth has a stabilizing effect. For large $m$ we find $\sigma_g^m = O(m^{1/2})$, which can readily be derived from the identity $\lambda_{m}^m = \Gamma(m+\frac{1}{2})\Gamma(1/2)/\Gamma(m+1)$ along with the asymptotic scaling $\Gamma(x+1)/\Gamma(x+s) \sim x^{1-s}$.

It is also interesting to compare this scaling to that of a colony growing on a rigid substrate (\ref{eq:u_rigid})-(\ref{eq:pbc_rigid}), in which the growth pressure acts locally. Linear stability analysis of this simpler model, which is included in appendix \ref{sec:rigid-stability}, gives $\sigma_{g,{\rm rigid}}^m = -(m - 1) / 2$. This is also linearly stable but has a linear scaling with respect to $m$. In this case we expect perturbations to be suppressed at a faster rate. Notably, the quantitative stabilizing behavior for growth on a viscous substrate is not captured by the local pressure model.

\subsection{Buoyancy}

We next analyze the stability of buoyancy dominated flow, ${\Ray}\to \infty$. Defining $\tbu = (\beta{\Ray}) \tbu_b$, $p = (\beta{\Ray}) p_b$, and $\sigma = \beta\sigma_b$ and taking ${\Ray}\to\infty$, the system of equations is
\begin{align}
    \tbu_b + J_{\feps}^{-1}\left(\tnabla \cS_{\feps}[p_b] + \frac{1}{16} \tnabla  \cB_{\feps}[\sigma_b]\right) = \bzero&,\\
    \cN_{\feps}[p_b] + \frac{1}{16} \cV_{\feps}[\sigma_b] = 0&,\\
    \sigma_b - 2 = 0&.
\end{align}
Similar to before, we expand
\begin{align}
\tbu_b &= R^2(\tbu_0 + \eps\tbu_1) + O(\eps^2),\\
p_b\circ\feps &= R^3(p_0 + \eps p_1) + O(\eps^2),\\
\sigma_b\circ\feps &= (\sigma_0 + \eps \sigma_1) + O(\eps^2). 
\end{align}
Substituting above and matching powers of $\eps$, the $O(1)$ equations are given by the unit disk equations
\begin{align}
    \tbu_0 + \tnabla\cS_D[p_0] + \frac{1}{16}\tnabla\cB_D[\sigma_0] &= \bzero,\\
    \cN_D[p_0] + \frac{1}{16} \cV_D[\sigma_0] &= 0,\\
    \sigma_0 - 2 &= 0,
\end{align}
and the $O(\eps)$ equations are
\begin{align}
\tbu_1 + \tnabla\left(\cS_D[p_1] + \cSeps{}[p_0]\right) + \frac{1}{16}\tnabla\left(\cB_D[\sigma_1] + \cBeps{}[\sigma_0]\right) &= \bzero,\label{eq:ueps_b}\\ 
\cN_D[p_1] + \cNeps{}[p_0] + \frac{1}{16}\left(\cV_D[\sigma_1] + \cVeps{}[\sigma_0]\right) &= 0, \label{eq:peps_b}\\
\sigma_1 &= 0.\label{eq:seps_b}
\end{align}
Note that we used $\tbu_0 + \tnabla(\cS_D[p_0] + (1/16)\cB_D[\sigma_0]) = \bzero$ from the $O(1)$ equation to simplify the first equation. In the following we assume $m \geq 1$; the case $m = 0$ is trivial.

The $O(1)$ solution $p_0$ is given by Eq. (\ref{eq:pb}) and we trivially have $\sigma_0 = 2$ and $\sigma_1 = 0$. In order to compute the perturbed pressure $p_1$, we must first evaluate $\cNeps{}[p_0]$ and $\cVeps{}[\sigma_0]$. Using identities (\ref{eq:N[wz^j]}), (\ref{eq:N[w^3z^j]}), and (\ref{eq:ilambda_m^m}), we find
\begin{align}
    \cNeps{}[p_0] =  \left[\frac{\lambda^0_1}{32(m+1)\lambda^m_{m+1}} \left(
    (m+2) - (2m+3)(1-|\zeta|^2)  
    \right) - \frac{1}{16} |\zeta|^2\right]\zeta^m.
\end{align}
Using the identity (\ref{eq:V_D[z^m]}) for $m \geq 1$, we have
\begin{equation}
    \cVeps{}[\sigma_0] = \left(|\zeta|^2 - \frac{1}{m}\right)\zeta^m.
\end{equation}
We can then use Eq. (\ref{eq:peps_b}) to solve for $p_1$,
\begin{align}
    p_1(\zeta) = \frac{\lambda^0_1}{96} \left(3 -  \frac{6\lambda^m_{m+1}}{m\lambda^0_1} - 4(1 -|\zeta|^2)
    \right)\omega(\zeta)\zeta^m.
\end{align}

To compute the velocity, we need to evaluate $\cS_D[p_1], \cSeps{}[p_0]$, and $\cBeps{}[\sigma_0]$. Using the identities (\ref{eq:S[wz^j]}), (\ref{eq:S[w^3z^j]}), and (\ref{eq:lambda_m^m}), we get
\begin{align}
\cS_D[p_1] &= \frac{3}{768}\left[\lambda^0_1
    \left(
    \lambda^{m+1}_{m+1} |\zeta|^2 - \lambda^{m+2}_{m+2} |\zeta|^4 
    \right) 
    - \frac{4}{m} \left( \frac{2}{2m+1} - \frac{|\zeta|^2}{m+1} \right)\right]\zeta^m, \\
    \cSeps{}[p_0] &=  \frac{1}{768} \left[\lambda^0_1\left(
 -2 \lambda^m_m
- 3 \lambda^{m+1}_{m+1} |\zeta|^2
+ 3 \lambda^{m+2}_{m+2} |\zeta|^4\right)
+ 12 \left(|\zeta|^2 - \frac{|\zeta|^4}{2}\right)
    \right] \zeta^m.
\end{align}
Further, using identity (\ref{eq:B_D[z^m]}), we find
\begin{equation}
    \delta B_D^1[\sigma_0] = \left[\frac{|\zeta|^4}{8} - \frac{5|\zeta|^2}{8} + \frac{1}{4}\right] \zeta
\end{equation}
and
\begin{equation}
    \cBeps{}[\sigma_0] = \left[\frac{|\zeta|^4}{8} - \frac{|\zeta|^2}{4}\left(1 + \frac{1}{m(m+1)}\right) + \frac{1}{4m(m-1)}\right]\zeta^m \quad {\rm for} ~ m \geq 2.
\end{equation}
For $m = 1$ this gives $\cS[p_1] + \cSeps{}[p_0] + (1/16)\cBeps{}[\sigma_0] = 0$ so that $\sigma_b^1 = 0$. Using Eq. (\ref{eq:ueps_b}), for $m \geq 2$ this yields the $O(\eps)$ interface velocity
\begin{equation}
\begin{aligned}
   (\tbu_1 \cdot \hat{\br})(e^{\i \theta}) &= -\Real \left[ \frac{\partial}{\partial r}\left(\cS_D[p_1] + \cSeps{}[p_0] + \frac{1}{16}\cBeps{}[\sigma_0]\right) \right ]\Big\rvert_{\zeta = e^{\i \theta}}\\
   &= \frac{1}{96}\left[m\frac{\lambda^0_1 \lambda^m_m}{4} - \frac{9}{2(m-1)(2m+1)}\right] \cos m\theta
   \\ & =: \sigma_b^m \cos m\theta.
\end{aligned}
\end{equation}

\begin{figure}[t]
    \centering
    \includegraphics[width=\linewidth]{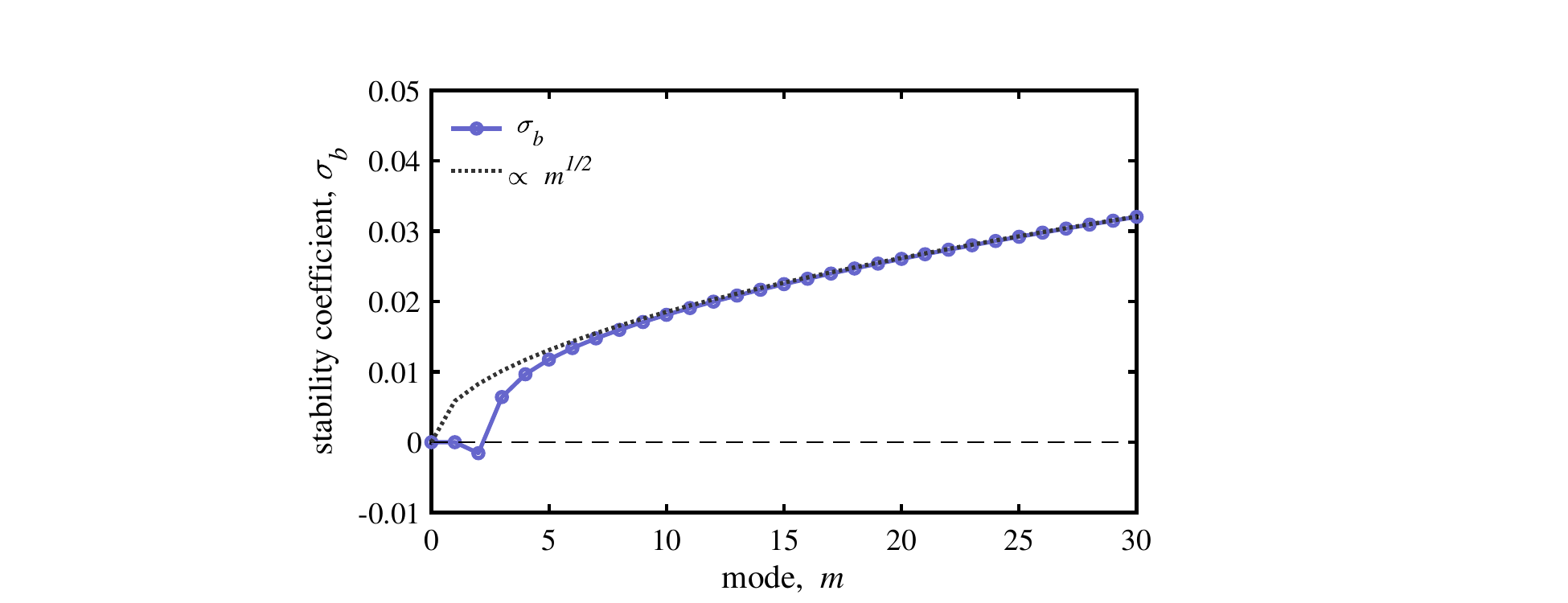}
    \caption{Stability coefficient $\sigma_b^m$ for buoyancy-dominated flow ($\Ray \to \infty$). For $m = 0, 1$, stability is marginal ($\sigma_b^0 = \sigma_b^1 = 0$), which is reflective of scale and translational invariance, respectively. All other coefficients are positive, indicating buoyancy-driven flows are destabilizing, and tend towards positive infinity as $O(m^{1/2})$.}
    \label{fig:sigma_b}
\end{figure}

A plot of $\sigma_b^m$ is shown in Fig. \ref{fig:sigma_b}. Both the $m = 0$ and $m = 1$ modes have $\sigma_b^m = 0$, which is consistent with scale and translational invariance. The $m = 2$ mode, which corresponds to an ellipsoidal-like perturbation, is negative, while all other modes are positive with an asymptotic scaling of $\sigma_b^m = O(m^{1/2})$ as in the growth-dominated case. This shows buoyancy-induced flows have a destabilizing effect.

\begin{figure}[t]
    \centering
    \includegraphics[width=\linewidth]{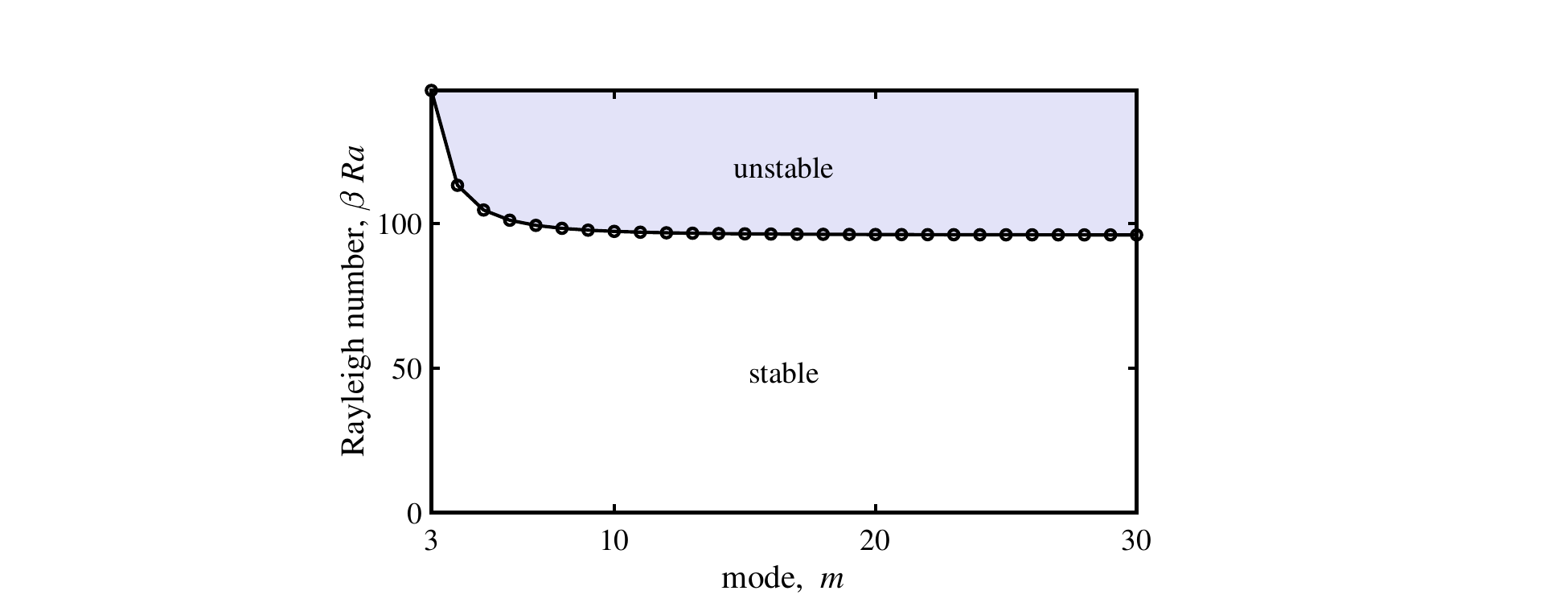}
    \caption{Stable (white) and unstable (purple) regions in the $m-\beta\Ray$ plane corresponding to the short time solution (\ref{eq:eps(t)-O(t)}). The black line and circle markers denote the stability threshold $\beta\Ray_*^m$ given by Eq. (\ref{eq:Ra_crit}). This critical value decreases with $m$, and asymptotically approaches a positive constant $\lim_{m\rightarrow\infty}\beta\Ray_*^m = 96$.}
    \label{fig:stability}
\end{figure}

\subsection{Stability criterion}\label{sec:stability-criterion}

\begin{figure}[t]
    \centering
    \includegraphics[width=\linewidth]{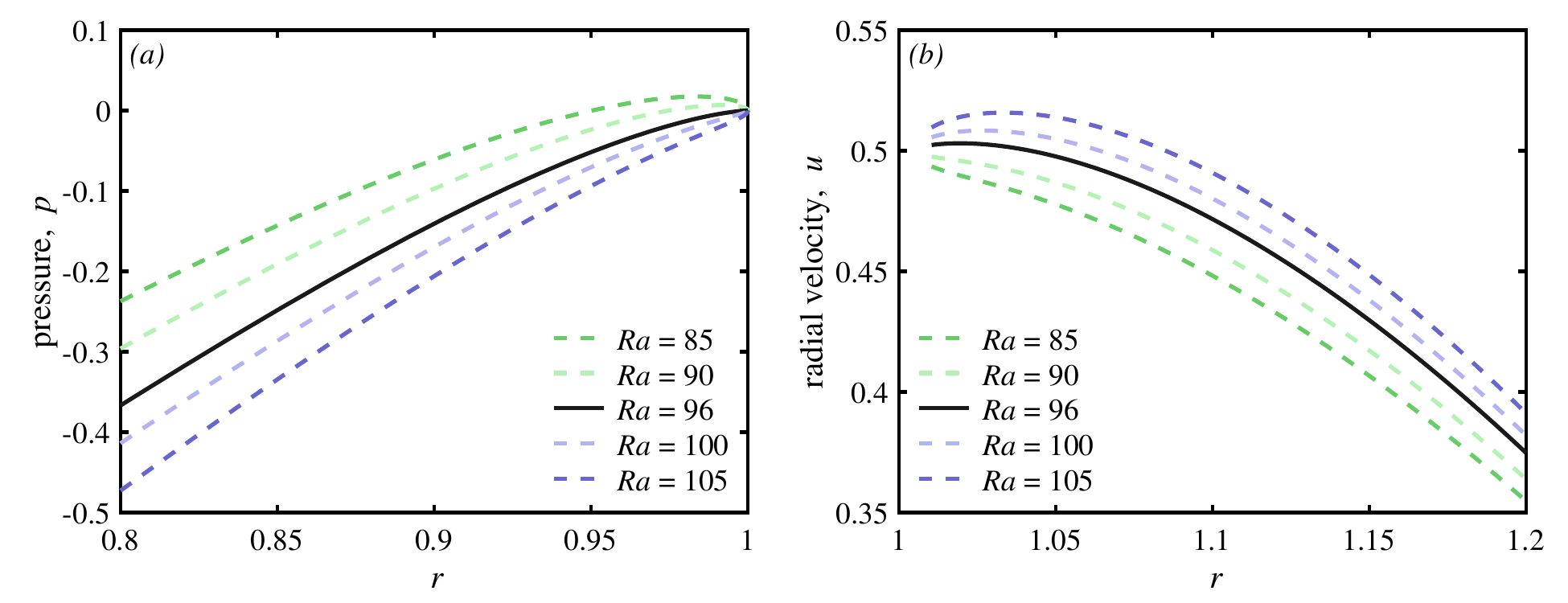}
    \caption{Axisymmetric {\it (a)} pressure and {\it (b)} velocity near the transition to instability $\Ray = 96$ (solid line) for $\beta = 1$. In the former, as $\Ray$ crosses the critical value, the pressure becomes negative everywhere and its normal derivative changes sign. In the latter, the normal derivative of the radial velocity changes sign and a local maximum appears outside the droplet.}
    \label{fig:transition}
\end{figure}

Combining the two solutions, the interface evolves as
\begin{equation}
    \frac{\partial\bGamma^m}{\partial t} = R\left(\frac{1}{2} + \eps (\sigma_g^m + R^2\beta{\Ray}\sigma_b^m) \cos m\theta\right)\hat\bn + O(\eps^2),\label{eq:dgamma/dt_m}
\end{equation}
where we emphasize $R, \eps$, and $\hat\bn$ depend on time. Differentiating the representation (\ref{eq:gamma_eps^m}) in time, using $\hat \bn = \hat\br + \eps(m + 1)\sin m\theta\hat\btheta + O(\eps^2)$, and matching powers of $\eps$, one can find a set of differential equations for $R$ and $\eps$. However, the normal and tangential components are incompatible. This can be resolved by adding an arbitrary tangential velocity $\eps T(\theta, t)\sin m \theta \hat\bn^\perp$ to the interface velocity which does not affect the geometry. Equation (\ref{eq:dgamma/dt_m}) then becomes
\begin{equation}
\begin{aligned}
\frac{\dot R}{R}\hat\br + \dot\eps \left( \cos m\theta \hat\br  + \sin m\theta \hat \btheta \right)
&= \frac{1}{2}\hat\br + \eps \Bigg[\left(-\frac{\dot R}{R} + \sigma_g^m + R^2 \beta \Ray \sigma_b^m\right) \cos m\theta \hat \br \\&+ \left(-\frac{\dot R}{R} + \frac{(m+1)}{2} + \frac{T(\theta, t)}{R} \right) \sin m \theta\hat \btheta \Bigg] + O(\eps^2),
\end{aligned}
\end{equation}
from which we see $T(\theta, t) = R\left[-(m+1)/2 + \sigma_g^m + R^2\Ray\sigma_b^m\right]$ ensures compatibility in the $\hat\br$ and $\hat\btheta$ components. Matching powers of $\eps$, we therefore have the coupled ODEs
\begin{align}
\dot R &= \frac{R}{2},\\
\dot\eps &= \left(-\frac{\dot R}{R} + \sigma_g^m + R^2\beta{\Ray} \sigma_b^m\right)\eps,
\end{align}
with initial conditions $R(0) = 1$ and $\eps(0) = \eps_0 \ll 1$. The $O(1)$ solution always evolves as $R(t) = e^{t/2}$, corresponding to the axisymmetric case. Substituting this into the $\eps$ equation, we have
\begin{equation}
    \frac{\d\eps}{\d t} = \left(-\frac{1}{2} + \sigma_g^m + e^t \beta{\Ray} \sigma_b^m \right)\eps.
\end{equation}
The solution to this equation is a combined exponential and double exponential,
\begin{equation}
    \eps(t) = \eps_0 e^{(\sigma_g^m-1/2)t + \beta{\Ray} \sigma_b^m (e^t-1)}.\label{eq:eps(t)}
\end{equation}
For small $t$ we may expand the upper exponential so that
\begin{equation}
    \eps(t) = \eps_0 e^{[(\sigma_g^m-1/2) + \beta{\Ray} \sigma_b^m]t} + O(t^2).\label{eq:eps(t)-O(t)}
\end{equation}
Thus, on short time scales, the condition for instability of mode $m$, for $m > 2$, is
\begin{equation}
     \Ray > {\Ray_*^m}  := \frac{1/2-\sigma_g^m}{\beta\sigma_b^m} = \frac{96}{\beta}\frac{1}{\left(1 - \frac{18}{(m-1)(2m+1)\lambda_1^0\lambda_m^m}\right)}.\label{eq:Ra_crit}
\end{equation}
A phase diagram depicting this stability threshold is shown in Fig. \ref{fig:stability}. The critical Rayleigh number decreases with $m$ and asymptotically approaches $\lim_{m\rightarrow \infty} \beta\Ray_*^m = 96$. As previously noted, this limit is the threshold for which the axisymmetric pressure is everywhere negative in $D$. From a physical interpretation, negative pressure acts to compress the droplet but is forbidden to by the divergence constraint. This qualitative change in the pressure, and correspondingly the velocity, can be seen in Fig. \ref{fig:transition}, where we also see a local maximum in the radial velocity outside the droplet at the transition point. 

We note that this stability condition is specifically for the constant $\kappa = 0$. The influence of this constant on stability can be found in appendix D, where we find variations in the critical Rayleigh number are of about one order of magnitude over the range of experimental conditions.

\section{Discussion}

We developed and analyzed a model for the growth of a microbial droplet suspended on the surface of a viscous fluid. The integro-differential formulation, which is posed solely on the droplet, naturally lead to axisymmetric solutions and allowed us to rigorously analyze their linear stability. Here we found growth stabilizes the axisymmetric solution at all scales while buoyancy-induced flows are destabilizing, with the smallest scale perturbations growing the fastest.

These results provide an analytical description of the onset of pattern formation observed in yeast colonies growing on viscous substrates \citep{atis2019microbial}. The stability threshold we identified, $\beta\Ray = 96$, is of the correct order of magnitude for the initial instability observed in experiments, which occurs at a viscosity around $\mu \approx ~ 450 {\rm Pa ~ s}$, or $\Ray \approx 50$ as estimated in section \ref{sec:dimensional-analysis}. Moreover, instability of high wavenumber perturbations implies the interface will be rough at the smallest scales, which is also consistent with the thin fingers that are observed. The predicted dependence on a single dimensionless parameter can be tested experimentally by varying the substrate viscosity, droplet size, and cell growth rate. These results also extend to a droplet confined between two semi-infinite fluid layers, assuming the interface is impermeable and there are no nutrients in the upper layer, with the viscosity replaced by the sum of the viscosities of the two layers (see appendix \ref{app:fourier}).

The integro-differential formulation of the model requires the domain to be semi-infinite while in reality experiments are done on relatively shallow domains. Finite depth effects are partially captured by the matching constant $\kappa$, but future work may directly compute the solution for finite size domains. This can be done using a similar approach to that taken by \cite{jia2022incompressible}, where the Fourier transform of the associated Green's function can be found analytically, or by using the free-space Green's function of the Boussinesq-Stokes system. Similar to analyses of monolayer domains \citep{lubensky1996hydrodynamics}, we expect a bottom wall will suppress perturbations when the underlying fluid layer becomes very thin. Moreover, the wall will likely eliminate the Stokes paradox-type divergence of the fluid velocity. 

There are several additional features of the experiments of \cite{atis2019microbial} that are not described by the present model. In particular, the substrate exhibits non-Newtonian rheology at high shear rates, which would require a modification of the bulk constitutive model. Viscoelastic relaxation of the fluid could also be considered, resulting in a time-dependent integral kernel at linear order. Another feature is hole formation and breakup of the droplet in the high metabolic Rayleigh number regime, which could be modeled by allowing the cell density to vary.

Finally, for large deformations of the droplet where linear theory breaks down, we must resort to numerical solution. This is a challenging task owing to both solving the free boundary problem, which requires tracking a highly corrugated interface, and evaluating singular integral operators with non-smooth densities \citep{bruno2013high, helsing2022solving, zhou2026modeling}. The conformal mapping approach used for the linear stability analysis is a first step in this direction, however adaptive schemes will likely be required when the conformal distortion becomes large. With a fully nonlinear numerical solver in hand, collective behaviors among disjoint domains, such as those that appear following the breakup of the droplet, could then be explored.

\begin{bmhead}[Acknowledgements]
We thank Severine Atis and Michael Shelley for useful discussions. The Flatiron Institute is a division of the Simons Foundation.
\end{bmhead}

\begin{bmhead}[Data Availability]
    The data that support the findings of this study are openly available at \url{https://github.com/scottweady/microbial-droplets}.
\end{bmhead}

\begin{bmhead}[Declaration of Interests]
    The authors report no conflict of interest.
\end{bmhead}

\begin{appen}

\section{Effects of line tension}\label{app:line-tension}

Similar to domain boundaries in Langmuir films, we expect the microbial droplet will have a line tension (the analog of surface tension for a two-dimensional domain), which typically acts to suppress high frequency perturbations  \citep{stone1995hydrodynamics,lubensky1996hydrodynamics,alexander2007domain}. This is described by the boundary condition
\begin{equation}
-p\hat{\mathbf{n}} = \lambda k \hat{\mathbf{n}} \quad {\rm on} ~ \partial\Omega,
\end{equation}
where $k$ is the curvature and $\lambda$ is the line tension constant. Choosing the characteristic length scale $\ell_c = R$, time scale $t_c = 1/\gamma_\infty$, and growth pressure scale $p_c = \mu \ell_c / t_c$ as in section \ref{sec:dimensional-analysis}, the boundary condition in dimensionless form is $-p' = \lambda'k'$, where $\lambda' = \lambda/(\mu \gamma_\infty R^2)$. Using the same estimates as in section \ref{sec:dimensional-analysis}, we find $1/(\mu\gamma_\infty R^2) \approx 7.7 \times 10^7$. The line tension constant $\lambda$ requires specific experimental measurements, however for chemical systems it is typically on the order of $1-100{\rm pN}$ \citep{alexander2007domain}. We therefore estimate $\lambda' = O(10^{-5}) - O(10^{-3})$, so that the boundary condition $p' = 0$ is plausible at low to moderate wave numbers, specifically $m \ll \lambda'^{-1/2}$.

\section{Fourier space solution}\label{app:fourier}

In section \ref{sec:BIE-Fourier} we found the system of ODEs for the Fourier coefficients of the velocity, pressure, and concentration,
\begin{align}
-\begin{pmatrix} \i\bk \\ \partial_z \end{pmatrix} \ft P + \left(-k^2 + \partial_z^2\right) \ft \bU - {\Ray} \ft c \be_z = \bzero,\\
\begin{pmatrix} \i\bk \\ \partial_z \end{pmatrix} \cdot \ft \bU = 0,\\
\left(-k^2 + \partial_z^2\right)\ft c = 0,
\end{align}
where $\bk \in \bbR^2$ is a wavevector with magnitude $k = |\bk|$. Let $\ft \bU(\bk, z) = (\ft \bu, \ft w)(\bk, z)$ and define $\ft \bu_0(\bk) = \ft\bu(\bk, 0)$, $\ft w_0(\bk) = \ft w(\bk, 0)$, and $\ft c_0(\bk) = \ft c(\bk, 0)$. The concentration can be determined independently, giving $\ft c(\bk, z) = c_0(\bk) e^{kz}$, where we choose the solution that decays as $z\rightarrow-\infty$. Further, the no-penetration boundary condition $\bU\cdot\be_z = 0$ implies $\ft w_0 = 0$. After some calculation, we then find
\begin{align}
\ft \bu(\bk, z) &= \left(\ft \bu_0(\bk) + \ft \bu_1(\bk) z + \ft \bu_2(\bk) z^2 \right) e^{k z}, \label{eq:u_hat}\\
\ft w(\bk, z) &= \left(\ft w_1(\bk) z + \ft w_2(\bk) z^2 \right) e^{k z},\label{eq:w_hat}\\
\ft P(\bk, z) &= \left( \ft P_0(\bk) + \ft P_1(\bk) z\right) e^{k z},
\end{align}
where the coefficients are
\begin{align}
\ft P_0(\bk) &= -2\left((\i \bk\cdot\ft\bu_0) + \frac{3{\Ray}\ft c_0}{8k}\right), \\
\ft P_1(\bk) &= -\frac{{\Ray} \ft c_0}{2},\\
\ft \bu_1(\bk) &= -\frac{\i\bk}{k}\left((\i \bk\cdot\ft\bu_0) + \frac{{\Ray}\ft c_0}{4k}\right),\\
\ft \bu_2(\bk) &= -\frac{\i\bk}{8k} {\Ray}\ft c_0,\\
\ft w_1(\bk) &= -(\i \bk\cdot\ft\bu_0),\\
\ft w_2(\bk) &= -\frac{{\Ray}\ft c_0}{8}.
\end{align}
Taking the Fourier transform of the stress boundary condition $T[\bfv]\rvert_{\Omega} = -\grads p$, we find
\begin{equation}
\begin{aligned}
    \frac{\partial \hat\bu}{\partial z}\Big\rvert_{z=0} &= -\int_\Omega e^{-\i\bk\cdot\bx} \grads p ~ \d\bx
    \\ & =
    +\int_\Omega (-\i\bk)e^{-\i\bk\cdot\bx} p ~ \d\bx - \int_{\partial\Omega} e^{-\i\bk\cdot\bGamma} p(\bGamma) ~ \d\bGamma
    \\ & = 
    -\i\bk \wft{\chi_\Omega p},
\end{aligned}
\end{equation}
where we used $p = 0$ on $\partial\Omega$ in the last equality. Differentiating equation (\ref{eq:u_hat}) in $z$ and evaluating at $z = 0$, we find
\begin{align}
\ft\bu_0 & = \i\bk\left( -\frac{\wft{\chi_\Omega p}}{2k} + \frac{\Ray}{8k^3} \ft c_0\right).
\end{align}

Note that this calculation is unchanged if there is an impermeable, nutrient-free fluid above the surface. In this case the (dimensional) boundary condition becomes $T[\mu^-\partial\bU^-/\partial z]\rvert_\Omega - T[\mu^+\partial\bU^+/\partial z]\rvert_\Omega = -\grads p$, where $\bU^-,\bU^+$ are the bulk velocities of the lower and upper fluids, respectively, and $\mu^-,\mu^+$ are the corresponding viscosities. In terms of $\hat\bu_0$, the solution in the lower layer is as before and the solution in the upper layer is analogous after taking $\Ray = 0$ and changing the sign of $w$ and $z$. We therefore find, in dimensionless form,
\begin{align}
(1 + \mu')\ft\bu_0 & = \i\bk\left( -\frac{\wft{\chi_\Omega p}}{2k} + \frac{\Ray}{8k^3} \ft c_0\right),
\end{align}
where $\mu' = \mu^+/\mu^-$ is the viscosity ratio.

\section{Derivation of the integral operators}\label{app:integral-operators}

The real-space velocity is given by the inverse Fourier transform of Eqs. (\ref{eq:u_hat})-(\ref{eq:w_hat}). Defining
\begin{align}
    I_n(\bx, z) = \cF^{-1}\left[k^ne^{kz}  \right](\bx, z)
\end{align}
and performing a bit of algebra, the three-dimensional velocity $\bU = (\bu, w)$ can be expressed as
\begin{align}
\bu(\bx, z) &= - \grads\left[\left(\frac{1}{2} I_{-1} + \frac{z}{2} I_0 \right)*(\chi_{\Omega} p)  + \frac{\Ray}{16} \left( I_{-4} - zI_{-3} - z^2 I_{-2} \right) * (\chi_{\Omega} \sigma)\right], \\
w(\bx, z) &=   -\left[ \frac{z}{2} I_1 * (\chi_{\Omega} p) + \frac{\Ray}{16} \left( zI_{-2} - z^2 I_{-1} \right) * (\chi_{\Omega} \sigma)\right],\label{eq:w-bulk}
\end{align}
where $*$ denotes convolution over $\bbR^2$. We are faced with the difficulty that for $n < -1$ the $I_n$ are not classically integrable and must be interpreted in a distributional sense. However, notice that $\partial_z^j I_n(\bx,z) = I_{n + j}(\bx,z)$ so that we only need to compute $I_{-4}$. We interpret this as the solution to the equation $\laps^2(I_{-4}) = \cF^{-1}[e^{kz}]$. The right hand side is
\begin{equation}
    \begin{aligned}
        \cF^{-1}[e^{kz}] &= \frac{1}{2\pi} \int_0^\infty k e^{kz} J_0(|\bx| k) ~ dk
        \\ & = -\frac{1}{2\pi} \frac{z}{(|\bx|^2 + z^2)^{3/2}},
    \end{aligned}
\end{equation}
where $J_\alpha$ is the Bessel function of the first kind. We therefore seek solutions in the variable $r = |\bx|$, which yields the general solution
\begin{equation}
\begin{aligned}
8\pi I_{-4}(r, z) &= c_1 r^2 \log r + c_2r^2 + c_3\log r + c_4
\\ & + \left(-3 z \sqrt{r^2 + z^2} + (z^2 - r^2) \log\left(\frac{r^2}{\sqrt{r^2 + z^2} - z}\right) - z^2\log\left(\sqrt{r^2 + z^2} - z\right)\right),
\end{aligned}
\end{equation}
where $c_1, c_2, c_3, c_4$ are integration constants which may depend on $z$. A necessary condition of axisymmetry is $\partial_r I_{-4} = \partial_r^3 I_{-4} = 0$, which implies $c_1 = 2$ and $c_3 = -2z^2$. It remains to determine $c_2$ and $c_4$. These two constants are related by the condition $(\laps + \partial_z^2) I_{-4} = 0$, which gives $c_4'' = -(r^2 c_2'' + 4c_2 + 4)$. For $c_4$ to be independent of $r$, we require $c_2(z) = Az + B$. This shows $c_4(z) = -(2A/3) z^3 - 2(B + 1)z^2 + Cz + D$. Because $\bu$ is a gradient in $\bx$, it can be shown that the constants $A, C, D$ do not affect the solution and can be set to zero without loss of generality. This leaves the single free constant $B$. We can simplify $I_{-4}$ to
\begin{align}
    I_{-4}(r,z) = \frac{1}{8 \pi} \left((r^2 - z^2)  \left(\log\left(\sqrt{r^2 + z^2} - z \right) + B \right)  - 3 z\sqrt{r^2 + z^2} - \left( \log\left(\sqrt{r^2 + z^2} - z \right) + (B+2)  \right)z^2 \right). 
\end{align}
The expressions for $I_{-3}, I_{-2}, I_{-1}, I_0, I_1$ can be recovered by differentiating this with respect to $z$, however the full expressions are cumbersome and we omit them here. For $z = 0$ the relevant operators are
\begin{align}
I_{-4}(r, 0) &= \frac{1}{8\pi} r^2 \left(\log r + B\right),\\
I_{-1}(r, 0) &= \frac{1}{2\pi} \frac{1}{r},
\end{align}
from which Eq. (\ref{eq:u}) follows. For analytical convenience we set $B = -\log R - 1 + \kappa$ for some constant $\kappa$.

\begin{figure}
    \centering
    \includegraphics[width=\linewidth]{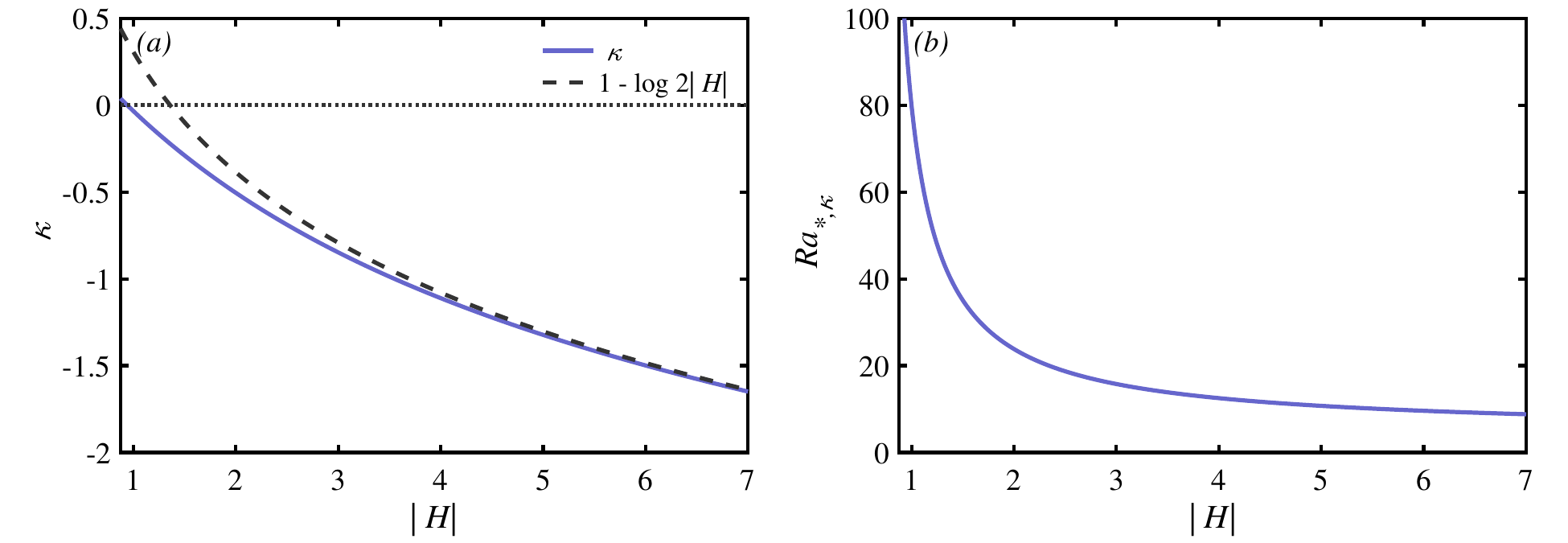}
    \caption{Dependence of stability on $\kappa$. Panel {\it (a)} shows $\kappa$ as a function of the effective domain height $H$ over the range of experimentally relevant values along with the asymptotic scaling $\kappa \sim 1 - \log 2|H|$. Panel {\it (b)} shows the corresponding critical Rayleigh number $\Ray_{*,\kappa}$ at $m\rightarrow\infty$ as a function of $H$. The critical value ranges over about one order of magnitude.}
    \label{fig:kappa}
\end{figure}

\section{Effect of the integration constant}\label{app:constant}

The constant $\kappa$ in the previous section relates to the far-field behavior of the velocity, which non-physically diverges when $\Ray > 0$. We interpret the integro-differential equation as an inner solution and $\kappa$ as a matching constant which is chosen such that the vertical velocity vanishes at a (dimensionless) domain depth $(r,z) = (0,H)$, and thus is small in the region $r/|H| \ll 1$. Figure \ref{fig:kappa} shows $\kappa$ as a function of $|H|$ over the experimental range $7/8 < |H| < 7$ reported in \cite{atis2019microbial}. An asymptotic analysis of the expression (\ref{eq:w-bulk}) shows $\kappa \sim 1 - \log 2|H|$, which gives a good characterization of the numerical solution even at small $|H|$. The choice $\kappa = 0$ used in the main analysis corresponds to an effective depth $|H| \approx 1$.

Now expanding the operator $\cB_{\Omega,\kappa}$ and using the notation $\cB_\Omega := \cB_{\Omega, 0}$, the velocity satisfies
\begin{equation}
    \bu + \nabla \cS_\Omega[p] + \frac{\Ray}{16}\cB_\Omega[\sigma] = -\frac{\Ray\kappa}{16} \nabla\left(\frac{1}{8\pi} \int_\Omega |\bx-\by|^2 \sigma(\by) ~ \d\by\right).
\end{equation}
We write $p = p' + p_\kappa$ where $p'$ solves the system with $\kappa=0$ and $p_\kappa$ satisfies
\begin{equation}
    \cN_\Omega[p_\kappa] = -\frac{\Ray\kappa}{16}\left(\frac{1}{2\pi}\int_\Omega \sigma(\by) ~ \d\by\right).
\end{equation}
For the nutrient rich case $\gamma(c) = 1$ for which $\sigma = 2\beta$, the right hand side is constant and we see $p_\kappa = (R^2\beta\Ray\kappa/16)p_g$. Following the same stability calculation, this yields
\begin{equation}
    \frac{\d\eps}{\d t} = \left[\left(1 + \frac{R^2\beta\Ray\kappa}{16}\right)\left(- \frac{1}{2} + \sigma_g^m\right) + R^2\beta\Ray\sigma_b^m \right]\eps.
\end{equation}
So long as $\kappa/16 < \sigma_b^m / (1/2-\sigma_g^m)$, the condition for instability of mode $m \geq 2$ becomes
\begin{align}
    \Ray > \Ray^m_{*,\kappa} := \frac{1}{\beta}\frac{\frac{1}{2} - \sigma_g^m}{\sigma_b^m + \frac{\kappa}{16}\left(-\frac{1}{2} + \sigma_g^m\right)} > 0.
\end{align}
First note that the pure growth case $\Ray = 0$ is unaffected by $\kappa$ and the axisymmetric solution is always stable. When $\Ray > 0$, the stability threshold depends on $\kappa$. In particular, because $(-1/2+\sigma_g^m) < 0$ when $m \geq 2$, in the limit $\kappa \rightarrow -\infty$, which corresponds to infinite depth, all modes are unstable, while for $\kappa > 1/6$ all modes are stable. Figure \ref{fig:kappa} shows $\Ray_{*,\kappa} := \lim_{m\rightarrow\infty} \Ray_{*,\kappa}^m$ as a function of $|H|$. We find $\Ray_{*,\kappa}$ ranges over about one order of magnitude for the experimental range $7/8 < |H| < 7$, and an instability is predicted in all cases.

\section{Recursion relations for the associated Legendre polynomials}\label{app:Plm_expansion}

Our calculations use several non-standard recursion relations of the associated Legendre polynomials. These are
\begin{align}
    P^\ell_{\ell+2}(x) &=  \frac{2\ell+1}{2}\left( -1 + (2\ell+3)x^2 \right)P^\ell_{\ell}(x), \\
    P^\ell_{\ell+3}(x) &=  \frac{2\ell+3}{2}\left( -1 + \left(\frac{2\ell+5}{3}\right) x^2 \right)P^\ell_{\ell+1}(x), \\
    P^\ell_{\ell+4}(x) &=  \frac{(2\ell+1)(2\ell+3)}{8}\left(1-2(2\ell+5)x^2 +  \frac{1}{3}(2\ell+5)(2\ell+7)x^4\right)P^\ell_{\ell}(x),
\end{align}
which also give the convenient identities
\begin{align}
    x^2 (1-x^2)^{\ell/2} &= \frac{(-1)^\ell}{(2\ell+3)!!} \left(
    (2\ell+1)P^\ell_{\ell}(x) + 2P^\ell_{\ell+2}(x)
    \right), \\
    x^3(1-x^2)^{\ell/2} &= 3\frac{(-1)^{\ell}}{(2\ell +5)!!} ((2\ell+3)P^{\ell}_{\ell+1}(x) + 2P^{\ell}_{\ell+3}(x)),\\
    x^4 (1-x^2)^{\ell/2} &=3 \frac{(-1)^\ell}{(2\ell+1)!!} \left(8P^\ell_{\ell+4}(x) +4(2\ell+5)P^\ell_{\ell+2}(x) +(2\ell+1)(2\ell+7)P^\ell_{\ell}(x)  \right).
\end{align}

\section{Evaluation of volume potentials}\label{app:volume_potentials}

In this section we evaluate the volume potentials $\cB_D[\zeta'^m]$ and $\cV_D[\zeta'^m]$. We will encounter integrals of the form
\begin{equation}
    A_{p,q}(z) = \int_D \log|z - \zeta| |\zeta|^p \zeta^q ~ \d\zeta.
\end{equation}
For $z \in \mathbb{C}\setminus D$ and $\zeta \in D$, we can make use of the expansion
\begin{equation}
    \log|z - \zeta| =
    \log|z| - \sum_{k=1}^\infty \frac{1}{2k}\left[\left(\frac{\zeta}{z}\right)^k + \left(\frac{\bar\zeta}{\bar z}\right)^k\right].\label{eq:log|z-zeta|}
\end{equation}
Using this expansion and orthogonality of the Fourier modes, we find
\begin{equation}
A_{p,q}(z) = \begin{cases}
-\frac{\pi}{(p+2)|q|} \frac{1}{z^{|q|}} & q < 0, \\
\frac{2\pi}{(p+2)} \log|z| & q = 0, \\
-\frac{\pi}{(p+2+2q)q} \frac{z^q}{|z|^{2q}} & q > 0,
\end{cases}
\quad{\rm for} ~ z \in \mathbb{C} \setminus D.
\label{eq:logpq}
\end{equation}

\subsection{Laplacian}

We have $\cV_D[\zeta'^m](z) = A_{0,m}(z)/2\pi$ and can apply Eq. (\ref{eq:logpq}) immediately,
\begin{equation}
    \cV_D[\zeta'^m](z) = 
    \begin{cases}
    \frac{\log|z|}{2} &\quad m = 0,\\
    -\frac{1}{4m(m+1)} \frac{z^m}{|z|^{2m}} &\quad m > 0,
    \end{cases}
    \quad{\rm for} ~ z \in \mathbb{C} \setminus D.
\end{equation}
Noting that $\Delta \cV_D[\zeta'^m] = \zeta^m$ in $D$, we have the general solution on the interior for all $m\geq0$,
\begin{equation}
    \cV_D[\zeta'^m] = (a_m|\zeta|^2 + b_m)\zeta^m \quad {\rm for} ~ \zeta \in D,\label{eq:V_D[z^m]}
\end{equation}
where $a_m = 1/(4(m+1))$. The coefficient $b_m$ can be determined by enforcing continuity of the solution across the boundary of $D$, which gives
\begin{equation}
    b_m = 
    \begin{cases}
        0 & m = 0, \\
        -\frac{1}{4m} & m \geq 1.
    \end{cases}
\end{equation}

\subsection{Bilaplacian}

Expanding $|z-\zeta|^2 = |z|^2 + |\zeta|^2 - z\bar\zeta - \bar z\zeta$, we have
\begin{equation}
    \cB_D[\zeta'^m](z) = \frac{1}{8\pi}\left[ |z|^2 A_{0,m} - zA_{2,m-1} - \bar z A_{0,m+1} + A_{2,m} - \left(\int_D |z-\zeta'|^2 \zeta'^m \d\zeta'\right)\right].
\end{equation}
We treat the $m = 0$ and $m = 1$ terms separately,
\begin{equation}
    \cB_D[1](z) = \frac{1}{8}\left( |z|^2\log|z| - |z|^2 + \frac{\log|z|}{2}\right) \quad {\rm for} ~ z \in \mathbb{C} \setminus D,
\end{equation}
and
\begin{equation}
    \cB_D[\zeta'](z) = \frac{1}{16}\left(-|z|^2\log|z| + \frac{|z|^2}{2}  -\frac{1}{6}\right)\frac{z}{|z|^2} \quad {\rm for} ~ z \in \mathbb{C} \setminus D.
\end{equation}
For $m \geq 2$, we get
\begin{equation}
    \cB_D[\zeta'^m](z) = \frac{1}{16m(m+1)}\left(\frac{|z|^2}{m-1} - \frac{1}{m + 2}\right)\frac{z^m}{|z|^{2m}} \quad {\rm for} ~ z \in \mathbb{C}\setminus D.
\end{equation}
Noting that $\Delta^2\cB_D[\zeta'^m] = \zeta^m$ in $D$, we have the general solution on the interior for all $m \geq 0$,
\begin{equation}
\cB_D[\zeta'^m] = (a_m|\zeta|^4 + b_m|\zeta|^2 + c_m)\zeta^m \quad {\rm for} ~ \zeta \in D,\label{eq:B_D[z^m]}
\end{equation}
where $a_m = 1/(32(m+1)(m+2))$. As before, the coefficients $b_m$ and $c_m$ can be determined by enforcing continuity of the solution and its radial derivative across the boundary of $D$, which gives
\begin{equation}
b_m =
\begin{cases}
-\frac{1}{16} & m = 0, \\
-\frac{1}{16m(m+1)} & m \geq 1,
\end{cases}
\end{equation}
and
\begin{equation}
c_m = 
\begin{cases}
-\frac{5}{64} & m = 0,\\
\frac{3}{64} & m = 1,\\
\frac{1}{32m(m-1)} & m \geq 2.
\end{cases}
\end{equation}

\section{Stability of the rigid substrate solution}\label{sec:rigid-stability}

In this section we perform a linear stability analysis of the rigid substrate problem
\begin{align}
\bu + \nabla p = \bzero &\quad {\rm in} ~ \Omega_\eps^m,\\
\nabla\cdot\bu = 1 &\quad {\rm in} ~ \Omega_\eps^m,\\
p = 0 &\quad {\rm on} ~ \partial\Omega_\eps^m,
\end{align}
where $\Omega_\eps^m$ is as in section \ref{sec:stability} (see e.g. \cite{weady2024mechanics} for further description of this model). Eliminating the velocity, we have
\begin{align}
-\Delta p = 1 &\quad {\rm in} ~ \Omega_\eps^m,\\
p = 0 &\quad {\rm on} ~ \partial\Omega_\eps^m.
\end{align}
Restating these equations on the unit disk in terms of the conformal map $\feps$, the pressure satisfies
\begin{align}
    -\tilde\Delta(p\circ\feps) = J_{\feps} &\quad{\rm in} ~ D,\\
    p\circ\feps = 0 &\quad{\rm on} ~ \partial D.
\end{align}
Making the expansion $p\circ\feps(\zeta) = R^2(p_0(\zeta) + \eps p_1(\zeta))$, we get the $O(1)$ equation
\begin{align}
    -\tilde\Delta p_0 = 1 &\quad {\rm in} ~ D,\\
    p_0 = 0 &\quad {\rm on} ~ \partial D,
\end{align}
and the $O(\eps)$ equation
\begin{align}
    -\tilde\Delta p_1 = 2(m + 1)\zeta^m &\quad {\rm in} ~ D,\\
    p_1 = 0 &\quad {\rm on} ~ \partial D.
\end{align}
It is readily checked that $p_0(\zeta) = (1 - |\zeta|^2)/4$ and $p_1(\zeta) = \zeta^m(1 - |\zeta|^2)/2$ satisfy the equations and boundary conditions. The normal interface velocity is therefore
\begin{equation}
    \begin{aligned}
        (\bu\cdot\hat\bn)(\theta) &= \Real\left[R(\tilde\bu\cdot\hat\br)\rvert_{e^{\i\theta}}\right]
        \\ & = 
        -R\left[\frac{\partial p_0}{\partial r} + \eps \Real\left(-(m + 1)\zeta^m\frac{\partial p_0}{\partial r} + \frac{\partial p_1}{\partial r}\right)\Big\rvert_{e^{\i\theta}}\right] + O(\eps^2)
        \\ & =
        R\left[\frac{1}{2} - \eps \left(\frac{m-1}{2}\right)\cos m\theta\right] + O(\eps^2).
    \end{aligned}
\end{equation}
Following the same approach as in section \ref{sec:stability} with initial data $R(0) = 1$ and $\eps(0) = \eps_0$, we have $R(t) = e^{t/2}$ and $\eps(t) = \eps_0 e^{-mt/2}$. Thus, the axisymmetric solution is stable to all perturbations and the damping rate is linear in $m$.

\section{Numerical validation}

In this section we verify our calculations through numerical solution of the $O(1)$ and $O(\eps)$ equations. The numerical methods used here also provide a baseline for solving the integro-differential equation under more general conditions that do not admit exact solutions. This requires a procedure for computing the operators $\cS_D, \cN_D$, $\cV_D$, and $\cB_D$. A central challenge is the singular kernel of the operators $\cS_D$ and $\cN_D$ and, moreover, the finite-part integral in the latter. Here we exploit the spectral properties of these operators to evaluate them numerically. 

\subsection{Eigenfunction expansions and quadrature}

Given a function $\nu(\bx)$, we wish to determine the (complex-valued) coefficients $\hat\nu_\ell^m$ such that
\begin{equation}
    \nu(\bx) = \sum_{\ell,m} \hat\nu_\ell^m y_\ell^m(\bx),
\end{equation}
where the sum is taken over even $\ell + m$ when $\nu \in C^\infty(D)$ and odd $\ell + m$ when $\nu / \omega \in C^\infty(D)$. In practice, we choose a maximal degree $M$ such that the eigenfunction expansion is truncated at $0\leq \ell \leq M$ and $|m| \leq \ell$. Using the orthogonality relation for the projected spherical harmonics, we have, for each admissible pair $(\ell, m)$,
\begin{equation}
    \hat\nu_\ell^m = \int_D \frac{y_\ell^{m*}(\bx)\nu(\bx)}{\omega(\bx)} ~ \d\bx.\label{eq:inner-product}
\end{equation}
To evaluate this integral, we make the change of coordinates $\bx(s, \theta) = \sqrt{1-s^2}(\cos\theta,\sin\theta)$ with $(s, \theta) \in [0,1]\times [0,2\pi)$, for which the integral becomes
\begin{equation}
    \hat\nu_\ell^m = \int_0^1 \int_0^{2\pi} y_\ell^{m*}(s, \theta) \nu(s, \theta) ~ \d s \d\theta.\label{eq:nu_ell^m}
\end{equation}
This double integral can be evaluated with spectral accuracy using Gauss-Legendre quadrature in $s$ along with the trapezoidal rule in $\theta$. In discrete form, given a function $f(\bx) = f(s, \theta)$ on $D$, this is given by
\begin{equation}
    \int_D f(\bx) ~ \d\bx \approx \sum_{i=0}^{N_s-1} \sum_{j=0}^{N_\theta-1} f(s_i, \theta_j) s_i w_i w_j,\label{eq:quadrature}
\end{equation}
where $s_i$ and $w_i$ are the Gauss-Legendre nodes and weights of order $N_s$, and $\theta_j = 2\pi j/N_\theta$ are equispaced points on the interval $[0, 2\pi)$ with corresponding weights $w_j = 2\pi/N_\theta$. Note that the factor $s_i$ cancels when computing the weighted inner product (\ref{eq:inner-product}). The number of quadrature nodes must be high enough to resolve the eigenfunctions. To do so, we choose $N_s = M + 1$ and $N_\theta = 2M + 1$.

\subsection{Numerical evaluation of $\cS_D$}\label{app:S_D}

Now suppose $\nu(\bx) = \tilde\nu(\bx)/\omega(\bx)$ where $\tilde\nu \in C^\infty(D)$ is smooth and $\omega(\bx) = \sqrt{1 - |\bx|^2}$ as before. We expand $\tilde\nu$ in the even basis of projected spherical harmonics so that
\begin{equation}
    \nu(\bx) = \frac{1}{\omega(\bx)}\left(\sum_{\ell + m ~ {\rm even}} \hat\nu_\ell^m y_\ell^m(\bx)\right),
\end{equation}
where $\hat\nu_\ell^m$ are approximated by Eqs. (\ref{eq:nu_ell^m})-(\ref{eq:quadrature}). Applying $\cS_D$, we find
\begin{equation}
    \cS_D[\nu](\bx) = \frac{1}{4}\left(\sum_{\ell+m ~ {\rm even}} \lambda_\ell^m \hat\nu_\ell^m y_\ell^m(\bx)\right).
\end{equation}
This also provides a straightforward method for solving a system of the form $\cS_D[\nu] = f$ with $f \in C^\infty(D)$. Namely, writing
\begin{equation}
    f(\bx) = \sum_{\ell + m ~ {\rm even}} \hat f_\ell^m y_\ell^m(\bx),
\end{equation}
we have
\begin{equation}
    \nu(\bx) = \frac{4}{\omega(\bx)}\left(\sum_{\ell + m ~ {\rm even}} \frac{\hat f_\ell^m}{\lambda_\ell^m} y_\ell^m(\bx)\right).
\end{equation}

\subsection{Numerical evaluation of $\cN_D$}\label{app:N_D}

The procedure for evaluating $\cN_D$ is analogous to that for $\cS_D$. Let $\nu(\bx)$ be such that $\nu(\bx) / \omega(\bx) \in C^\infty(D)$. We expand $\nu$ in the odd basis of projected spherical harmonics so that
\begin{equation}
    \nu(\bx) = \sum_{\ell + m ~ {\rm odd}} \hat\nu_\ell^m y_\ell^m(\bx),
\end{equation}
where $\hat\nu_\ell^m$ are approximated by Eqs. (\ref{eq:nu_ell^m})-(\ref{eq:quadrature}). Applying $\cN_D$, we find
\begin{equation}
    \cN_D[\nu](\bx) = -\frac{1}{\omega(\bx)}\left(\sum_{\ell+m ~ {\rm odd}} \frac{1}{\lambda_\ell^m} \hat\nu_\ell^m y_\ell^m(\bx)\right).
\end{equation}
As before, this provides a straightforward method for solving a system of the form $\cN_D[\nu] = f$ with $f \in C^\infty(D)$. Namely, writing
\begin{equation}
    f(\bx) = \frac{1}{\omega(\bx)}\left(\sum_{\ell + m ~ {\rm odd}} \hat f_\ell^m y_\ell^m(\bx)\right),
\end{equation}
we have
\begin{equation}
    \nu(\bx) = -\left(\sum_{\ell + m ~ {\rm odd}} \lambda_\ell^m\hat f_\ell^m y_\ell^m(\bx)\right).
\end{equation}

\subsection{Numerical evaluation of $\cV_D$}\label{app:V_D}

We need to compute integrals of the form
\begin{equation}
    \cV_D[\nu](\bx) = \frac{1}{2\pi} \int_D \log|\bx-\bx'| \nu(\bx') ~ \d\bx'.    
\end{equation}
This can be accurately evaluated using singularity subtraction. Specifically, we decompose
\begin{equation}
\begin{aligned}
    \cV_D[\nu](\bx) &= -\frac{1}{2\pi} \int_D \log|\bx-\bx'| (\nu(\bx) - \nu(\bx')) ~ \d\bx' + \left(\frac{1}{2\pi} \int_D \log|\bx-\bx'| ~ \d\bx'\right) \nu(\bx)
    \\ & = 
    -\frac{1}{2\pi} \int_D \log|\bx-\bx'| (\nu(\bx) - \nu(\bx')) ~ \d\bx' + \left(\frac{|\bx|^2 - 1}{4}\right) \nu(\bx).
\end{aligned}
\end{equation}
If $\nu$ is H\"older continuous on $D$ we have $\lim_{\bx'\rightarrow\bx}\log|\bx-\bx'|(\nu(\bx)-\nu(\bx')) = 0$ and the first integral can be treated with the smooth quadrature rule (\ref{eq:quadrature}). 

\subsection{Numerical evaluation of $\cB_D$}\label{app:B_D}

We need to compute integrals of the form
\begin{equation}
    \cB_D[\nu](\bx) = \frac{1}{8\pi}\int_D |\bx-\bx'|^2(\log|\bx-\bx'| - 1) \nu(\bx') ~\d\bx'.
\end{equation}
If $\nu$ is bounded on $D$ the integrand vanishes for $\bx = \bx'$ and this integral can be directly evaluated using the quadrature rule (\ref{eq:quadrature}).

\vspace{0.25in}

Figure \ref{fig:convergence}{\em (a)} compares the numerical solution for the stability coefficient to the exact solution for perturbations up to $m = 8$. The convergence study in panel {\em (b)} shows the growth stability coefficient $\sigma_g^m$ is resolved to near machine precision for all $M$, which is consistent with the fact that the solution is exactly representable in the projected spherical harmonic basis. The buoyancy stability coefficient appears to converge as $O(M^{-4})$, achieving approximately eight digits of accuracy for an expansion of degree $M = 128$.

\begin{figure}
    \centering
    \includegraphics[width=\linewidth]{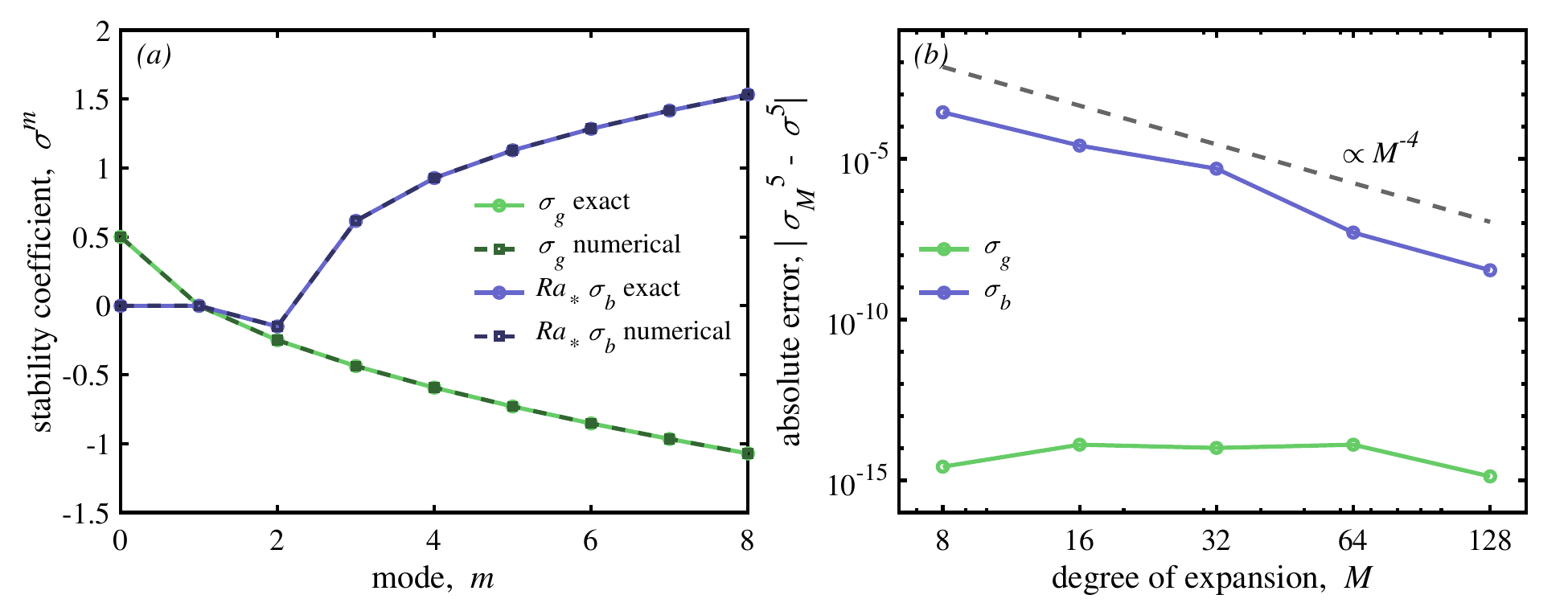}
    \caption{{\it (a)} Comparison between the exact and numerical stability coefficients for $M = 32$. The two are essentially indistinguishable across all modes tested. {\it (b)} A refinement study shows the numerical solution for the buoyancy coefficient $\sigma_b^m$ converges at roughly $O(M^{-4})$ order while $\sigma_g^m$ is resolved to near machine precision.}\label{fig:convergence}
\end{figure}

\end{appen}

\bibliographystyle{jfm}
\bibliography{refs}

@article{matar2009dynamics,
  title={Dynamics of surfactant-assisted spreading},
  author={Matar, Omar K and Craster, Richard V},
  journal={Soft Matter},
  volume={5},
  number={20},
  pages={3801--3809},
  year={2009},
  publisher={Royal Society of Chemistry}
}

@article{troian1990model,
  title={Model for the fingering instability of spreading surfactant drops},
  author={Troian, SM and Herbolzheimer, E and Safran, SA},
  journal={Physical review letters},
  volume={65},
  number={3},
  pages={333},
  year={1990},
  publisher={APS}
}

@article{hughes1981translational,
  title={The translational and rotational drag on a cylinder moving in a membrane},
  author={Hughes, BD and Pailthorpe, BA and White, LR},
  journal={Journal of Fluid Mechanics},
  volume={110},
  pages={349--372},
  year={1981},
  publisher={Cambridge University Press}
}

@article{zhou2026modeling,
  title={Modeling and Simulation of Open Membranes in Stokes Flow with Mixed-Dimensional Coupling},
  author={Zhou, Han and Young, Yuan-Nan and Mori, Yoichiro},
  journal={Multiscale Modeling \& Simulation},
  volume={24},
  number={2},
  pages={474--500},
  year={2026},
  publisher={SIAM}
}

@book{pozrikidis1992boundary,
  title={Boundary integral and singularity methods for linearized viscous flow},
  author={Pozrikidis, Constantine},
  year={1992},
  publisher={Cambridge university press}
}

@article{schneider1973slow,
  title={On the slow viscous rotation of a body straddling the interface between two immiscible semi-infinite fluids},
  author={Schneider, Joanne C and O'Neill, Michael E and Brenner, Howard},
  journal={Mathematika},
  volume={20},
  number={2},
  pages={175--196},
  year={1973},
  publisher={London Mathematical Society}
}

@article{stone2015mobility,
  title={Mobility of membrane-trapped particles},
  author={Stone, Howard A and Masoud, Hassan},
  journal={Journal of Fluid Mechanics},
  volume={781},
  pages={494--505},
  year={2015},
  publisher={Cambridge University Press}
}

@article{elfring2016surface,
  title={Surface viscosity and Marangoni stresses at surfactant laden interfaces},
  author={Elfring, Gwynn J and Leal, L Gary and Squires, Todd M},
  journal={Journal of Fluid Mechanics},
  volume={792},
  pages={712--739},
  year={2016},
  publisher={Cambridge University Press}
}

@article{masoud2014reciprocal,
  title={A reciprocal theorem for Marangoni propulsion},
  author={Masoud, Hassan and Stone, Howard A},
  journal={Journal of Fluid Mechanics},
  volume={741},
  pages={R4},
  year={2014},
  publisher={Cambridge University Press}
}

@article{stone1995hydrodynamics,
  title={Hydrodynamics of quantized shape transitions of lipid domains},
  author={Stone, HA and McConnell, HM},
  journal={Proceedings of the Royal Society of London. Series A: Mathematical and Physical Sciences},
  volume={448},
  number={1932},
  pages={97--111},
  year={1995},
  publisher={The Royal Society London}
}

@article{lubensky1996hydrodynamics,
  title={Hydrodynamics of monolayer domains at the air--water interface},
  author={Lubensky, David K and Goldstein, Raymond E},
  journal={Physics of fluids},
  volume={8},
  number={4},
  pages={843--854},
  year={1996},
  publisher={American Institute of Physics}
}

@article{alexander2007domain,
  title={Domain relaxation in {Langmuir} films},
  author={Alexander, James C and Bernoff, Andrew J and Mann, Elizabeth K and Mann, J Adin and Wintersmith, Jacob R and Zou, Lu},
  journal={Journal of Fluid Mechanics},
  volume={571},
  pages={191--219},
  year={2007},
  publisher={Cambridge University Press}
}

@article{stone1998hydrodynamics,
  title={Hydrodynamics of particles embedded in a flat surfactant layer overlying a subphase of finite depth},
  author={Stone, Howard A and Ajdari, Armand},
  journal={Journal of Fluid Mechanics},
  volume={369},
  pages={151--173},
  year={1998},
  publisher={Cambridge University Press}
}

@article{saffman1975brownian,
  title={Brownian motion in biological membranes.},
  author={Saffman, PG and Delbr{\"u}ck, M},
  journal={Proceedings of the National Academy of Sciences},
  volume={72},
  number={8},
  pages={3111--3113},
  year={1975}
}

@article{saffman1976brownian,
  title={Brownian motion in thin sheets of viscous fluid},
  author={Saffman, PG},
  journal={Journal of Fluid Mechanics},
  volume={73},
  number={4},
  pages={593--602},
  year={1976},
  publisher={Cambridge University Press}
}

@article{perlekar2010population,
  title={Population dynamics at high {Reynolds} number},
  author={Perlekar, P and Benzi, R and Nelson, DR and Toschi, F},
  journal={Physical Review Letters},
  volume={105},
  number={14},
  pages={144501--144501},
  year={2010}
}

@article{vaccari2015films,
  title={Films of bacteria at interfaces: three stages of behaviour},
  author={Vaccari, Liana and Allan, Daniel B and Sharifi-Mood, Nima and Singh, Aayush R and Leheny, Robert L and Stebe, Kathleen J},
  journal={Soft Matter},
  volume={11},
  number={30},
  pages={6062--6074},
  year={2015},
  publisher={Royal Society of Chemistry}
}

@article{hickl2022tubulation,
  title={Tubulation and dispersion of oil by bacterial growth on droplets},
  author={Hickl, Vincent and Juarez, Gabriel},
  journal={Soft Matter},
  volume={18},
  number={37},
  pages={7217--7228},
  year={2022},
  publisher={Royal Society of Chemistry}
}

@article{prasad2023alcanivorax,
  title={Alcanivorax borkumensis biofilms enhance oil degradation by interfacial tubulation},
  author={Prasad, M and Obana, N and Lin, S-Z and Zhao, S and Sakai, K and Blanch-Mercader, C and Prost, J and Nomura, N and Rupprecht, J-F and Fattaccioli, J and others},
  journal={Science},
  volume={381},
  number={6659},
  pages={748--753},
  year={2023},
  publisher={American Association for the Advancement of Science}
}

@article{aung2024comprehensive,
  title={A comprehensive review on kombucha biofilms: A promising candidate for sustainable food product development},
  author={Aung, Thinzar and Kim, Mi Jeong},
  journal={Trends in Food Science \& Technology},
  volume={144},
  pages={104325},
  year={2024},
  publisher={Elsevier}
}

@article{vaccari2017films,
  title={Films of bacteria at interfaces},
  author={Vaccari, Liana and Molaei, Mehdi and Niepa, Tagbo HR and Lee, Daeyeon and Leheny, Robert L and Stebe, Kathleen J},
  journal={Advances in colloid and interface science},
  volume={247},
  pages={561--572},
  year={2017},
  publisher={Elsevier}
}

@article{chang2015biofilm,
  title={Biofilm formation in geometries with different surface curvature and oxygen availability},
  author={Chang, Ya-Wen and Fragkopoulos, Alexandros A and Marquez, Samantha M and Kim, Harold D and Angelini, Thomas E and Fern{\'a}ndez-Nieves, Alberto},
  journal={New Journal of Physics},
  volume={17},
  number={3},
  pages={033017},
  year={2015},
  publisher={IOP Publishing}
}

@article{postek2024substrate,
  title={Substrate geometry affects population dynamics in a bacterial biofilm},
  author={Postek, Witold and Sta{\'s}kiewicz, Klaudia and Lilja, Elin and Wac{\l}aw, Bart{\l}omiej},
  journal={Proceedings of the National Academy of Sciences},
  volume={121},
  number={17},
  pages={e2315361121},
  year={2024},
  publisher={National Academy of Sciences}
}

@article{gu2016escherichia,
  title={How {Escherichia} coli lands and forms cell clusters on a surface: a new role of surface topography},
  author={Gu, Huan and Chen, Aaron and Song, Xinran and Brasch, Megan E and Henderson, James H and Ren, Dacheng},
  journal={Scientific reports},
  volume={6},
  number={1},
  pages={29516},
  year={2016},
  publisher={Nature Publishing Group UK London}
}

@article{gonzalez2025morphogenesis,
  title={Morphogenesis of bacterial cables in polymeric environments},
  author={Gonzalez La Corte, Sebastian and Stevens, Corey A and C{\'a}rcamo-Oyarce, Gerardo and Ribbeck, Katharina and Wingreen, Ned S and Datta, Sujit S},
  journal={Science Advances},
  volume={11},
  number={3},
  pages={eadq7797},
  year={2025},
  publisher={American Association for the Advancement of Science}
}

@article{asp2022spreading,
  title={Spreading rates of bacterial colonies depend on substrate stiffness and permeability},
  author={Asp, Merrill E and Ho Thanh, Minh-Tri and Germann, Danielle A and Carroll, Robert J and Franceski, Alana and Welch, Roy D and Gopinath, Arvind and Patteson, Alison E},
  journal={PNAS nexus},
  volume={1},
  number={1},
  pages={pgac025},
  year={2022},
  publisher={Oxford University Press}
}

@article{faiza2025substrate,
  title={Substrate stiffness modulates collective colony expansion of the social bacterium {Myxococcus} xanthus},
  author={Faiza, Nuzhat and Welch, Roy and Patteson, Alison},
  journal={APL bioengineering},
  volume={9},
  number={1},
  year={2025},
  publisher={AIP Publishing}
}

@article{fei2020nonuniform,
  title={Nonuniform growth and surface friction determine bacterial biofilm morphology on soft substrates},
  author={Fei, Chenyi and Mao, Sheng and Yan, Jing and Alert, Ricard and Stone, Howard A and Bassler, Bonnie L and Wingreen, Ned S and Ko{\v{s}}mrlj, Andrej},
  journal={Proceedings of the National Academy of Sciences},
  volume={117},
  number={14},
  pages={7622--7632},
  year={2020},
  publisher={National Academy of Sciences}
}

@article{abraham1998generation,
  title={The generation of plankton patchiness by turbulent stirring},
  author={Abraham, Edward R},
  journal={Nature},
  volume={391},
  number={6667},
  pages={577--580},
  year={1998},
  publisher={Nature Publishing Group UK London}
}

@article{hallegraeff2003harmful,
  title={Harmful algal blooms: a global overview},
  author={Hallegraeff, Gustaaf M},
  journal={Manual on harmful marine microalgae},
  volume={33},
  pages={1--22},
  year={2003},
  publisher={Unesco Paris}
}

@article{pearce2019flow,
  title={Flow-induced symmetry breaking in growing bacterial biofilms},
  author={Pearce, Philip and Song, Boya and Skinner, Dominic J and Mok, Rachel and Hartmann, Raimo and Singh, Praveen K and Jeckel, Hannah and Oishi, Jeffrey S and Drescher, Knut and Dunkel, J{\"o}rn},
  journal={Physical review letters},
  volume={123},
  number={25},
  pages={258101},
  year={2019},
  publisher={APS}
}

@article{greenspan1976growth,
  title={On the growth and stability of cell cultures and solid tumors},
  author={Greenspan, Harvey P},
  journal={Journal of theoretical biology},
  volume={56},
  number={1},
  pages={229--242},
  year={1976},
  publisher={Elsevier}
}

@article{lowengrub2009nonlinear,
  title={Nonlinear modelling of cancer: bridging the gap between cells and tumours},
  author={Lowengrub, John S and Frieboes, Hermann B and Jin, Fang and Chuang, Yao-Li and Li, Xiangrong and Macklin, Paul and Wise, Steven M and Cristini, Vittorio},
  journal={Nonlinearity},
  volume={23},
  number={1},
  pages={R1},
  year={2009},
  publisher={IOP Publishing}
}

@article{askham2025integral,
  title={Integral equations for flexural-gravity waves: analysis and numerical methods},
  author={Askham, Travis and Hoskins, Jeremy G and Nekrasov, Peter and Rachh, Manas},
  journal={arXiv preprint arXiv:2501.00887},
  year={2025}
}

@article{williams2004oblique,
  title={Oblique scattering of plane flexural--gravity waves by heterogeneities in sea--ice},
  author={Williams, Timothy D and Squire, Vernon A},
  journal={Proceedings of the Royal Society of London. Series A: Mathematical, Physical and Engineering Sciences},
  volume={460},
  number={2052},
  pages={3469--3497},
  year={2004},
  publisher={The Royal Society}
}

@article{chakrabarti2000solution,
  title={On the solution of the problem of scattering of surface--water waves by the edge of an ice cover},
  author={Chakrabarti, A},
  journal={Proceedings of the Royal Society of London. Series A: Mathematical, Physical and Engineering Sciences},
  volume={456},
  number={1997},
  pages={1087--1099},
  year={2000},
  publisher={The Royal Society}
}

@article{narayanasamy2025metabolically,
  title={Metabolically driven flows enable exponential growth in macroscopic multicellular yeast},
  author={Narayanasamy, Nishant and Bingham, Emma and Fadero, Tanner and Bozdag, G Ozan and Ratcliff, William C and Yunker, Peter and Thutupalli, Shashi},
  journal={Science Advances},
  volume={11},
  number={25},
  pages={eadr6399},
  year={2025},
  publisher={American Association for the Advancement of Science}
}

@article{weady2024mechanics,
  title={Mechanics and morphology of proliferating cell collectives with self-inhibiting growth},
  author={Weady, Scott and Palmer, Bryce and Lamson, Adam and Kim, Taeyoon and Farhadifar, Reza and Shelley, Michael J},
  journal={Physical Review Letters},
  volume={133},
  number={15},
  pages={158402},
  year={2024},
  publisher={APS}
}

@article{hallatschek2023proliferating,
  title={Proliferating active matter},
  author={Hallatschek, Oskar and Datta, Sujit S and Drescher, Knut and Dunkel, J{\"o}rn and Elgeti, Jens and Waclaw, Bartek and Wingreen, Ned S},
  journal={Nature Reviews Physics},
  volume={5},
  number={7},
  pages={407--419},
  year={2023},
  publisher={Nature Publishing Group UK London}
}

@article{benzi2022spatial,
  title={Spatial population genetics with fluid flow},
  author={Benzi, Roberto and Nelson, David R and Shankar, Suraj and Toschi, Federico and Zhu, Xiaojue},
  journal={Reports on Progress in Physics},
  volume={85},
  number={9},
  pages={096601},
  year={2022},
  publisher={IOP Publishing}
}

@article{jia2022incompressible,
  title={Incompressible active phases at an interface. Part 1. Formulation and axisymmetric odd flows},
  author={Jia, Leroy L and Irvine, William TM and Shelley, Michael J},
  journal={Journal of Fluid Mechanics},
  volume={951},
  pages={A36},
  year={2022},
  publisher={Cambridge University Press}
}

@article{manikantan2020surfactant,
  title={Surfactant dynamics: hidden variables controlling fluid flows},
  author={Manikantan, Harishankar and Squires, Todd M},
  journal={Journal of fluid mechanics},
  volume={892},
  pages={P1},
  year={2020},
  publisher={Cambridge University Press}
}

@article{atis2019microbial,
  title={Microbial range expansions on liquid substrates},
  author={Atis, Severine and Weinstein, Bryan T and Murray, Andrew W and Nelson, David R},
  journal={Physical review X},
  volume={9},
  number={2},
  pages={021058},
  year={2019},
  publisher={APS}
}

@article{fei2017active,
  title={Active colloidal particles at fluid-fluid interfaces},
  author={Fei, Wenjie and Gu, Yang and Bishop, Kyle JM},
  journal={Current opinion in colloid \& interface science},
  volume={32},
  pages={57--68},
  year={2017},
  publisher={Elsevier}
}

@article{masoud2014collective,
  title={Collective surfing of chemically active particles},
  author={Masoud, Hassan and Shelley, Michael J},
  journal={Physical review letters},
  volume={112},
  number={12},
  pages={128304},
  year={2014},
  publisher={APS}
}

@article{costabel2003asymptotics,
author = {Costabel, Martin and Dauge, Monique and Duduchava, Roland},
year = {2003},
month = {01},
pages = {869-926},
title = {Asymptotics Without Logarithmic Terms for Crack Problems},
volume = {28},
journal = {Communications in Partial Differential Equations},
doi = {10.1081/PDE-120021180}
}

@article{stephan1987boundary,
  title={Boundary integral equations for screen problems in $\mathbb{R}^3$},
  author={Stephan, Ernst P},
  journal={Integral Equations and Operator Theory},
  volume={10},
  number={2},
  pages={236--257},
  year={1987},
  publisher={Springer}
}

@article{wolfe1971eigenfunctions,
  title={Eigenfunctions of the integral equation for the potential of the charged disk},
  author={Wolfe, Peter},
  journal={Journal of Mathematical Physics},
  volume={12},
  number={7},
  pages={1215--1218},
  year={1971},
  publisher={American Institute of Physics}
}

@article{boersma1993solution,
  title={On the solution of an integral equation arising in potential problems for circular and elliptic disks},
  author={Boersma, J and Danick, E},
  journal={SIAM Journal on Applied Mathematics},
  volume={53},
  number={4},
  pages={931--941},
  year={1993},
  publisher={SIAM}
}

@article{martin1996mapping,
  title={Mapping flat cracks onto penny-shaped cracks, with application to somewhat circular tensile cracks},
  author={Martin, PA},
  journal={Quarterly of Applied Mathematics},
  volume={54},
  number={4},
  pages={663--675},
  year={1996}
}

@article{bruno2013high,
  title={A high-order integral solver for scalar problems of diffraction by screens and apertures in three-dimensional space},
  author={Bruno, Oscar P and Lintner, St{\'e}phane K},
  journal={Journal of Computational Physics},
  volume={252},
  pages={250--274},
  year={2013},
  publisher={Elsevier}
}

@article{helsing2022solving,
  title={Solving {Fredholm} second-kind integral equations with singular right-hand sides on non-smooth boundaries},
  author={Helsing, Johan and Jiang, Shidong},
  journal={Journal of Computational Physics},
  volume={448},
  pages={110714},
  year={2022},
  publisher={Elsevier}
}

\end{document}